\newcommand{\x}{{\bm x}}
\newcommand{\f}{{\bm f}}
\newcommand{\y}{{\bm y}}
\newcommand{\g}{{\bm g}}
\newcommand{\uu}{{\bm u}}

\newcommand{\cA}{{\mathcal{A}}}
\newcommand{\cF}{{\mathcal{F}}}

\newcommand{\res}{{\rm Res}}

\documentclass[times,1p]{elsarticle}
\usepackage{time,pslatex}
\usepackage{float}
\usepackage{algorithm}
\usepackage{algpseudocode}

\usepackage{epsfig}

\usepackage{amssymb}
 \usepackage{amsthm}
 \usepackage{amsmath}
 \usepackage{algpseudocode}
   \usepackage[dvipsnames]{xcolor}
\usepackage[colorlinks, linkcolor=reference,
            citecolor=citation,
          urlcolor=e-mail]{hyperref}

\definecolor{e-mail}{rgb}{0,.40,.80}
\definecolor{reference}{rgb}{.20,.60,.22}
\definecolor{citation}{rgb}{0,.40,.80}

\theoremstyle{definition}
\newtheorem{theorem}{Theorem}[section]

\newtheorem{lemma}[theorem]{Lemma}

\theoremstyle{definition}
\newtheorem{example}{Example}
\newtheorem{remark}{Remark}
\newtheorem{definition}{Definition}

\DeclareMathOperator{\rank}{rank}
\DeclareMathOperator{\supp}{supp}
\DeclareMathOperator{\proj}{proj}
\DeclareMathOperator{\diam}{diam}

\usepackage[numbers]{natbib}

\RequirePackage{color}
\definecolor{todo}{rgb}{1,0,0}
\definecolor{answer}{rgb}{0,0,1}
\definecolor{new}{rgb}{1,0,1}
\definecolor{conditional}{rgb}{0,1,0}
\definecolor{e-mail}{rgb}{0,.40,.80}
\definecolor{reference}{rgb}{.20,.60,.22}
\definecolor{mrnumber}{rgb}{.80,.40,0}
\definecolor{citation}{rgb}{0,.40,.80}

\journal{Journal of Symbolic Computation}

\usepackage[colorlinks, linkcolor=reference,
 citecolor=citation,
         urlcolor=e-mail]{hyperref}
\usepackage{url}     
\usepackage{bm}

\begin{document}

\begin{frontmatter}

\title{Symbolic-numeric algorithm for parameter estimation in discrete-time models with $\exp$\tnoteref{t1}}
\tnotetext[t1]{Dedicated to the memory of Marko Petkov\v{s}ek}

\author[add2]{Yosef Berman}
\ead{yberman1@gradcenter.cuny.edu}

\author[add6]{Joshua Forrest}
\ead{jeforrest@student.unimelb.edu.au}

\author[add2]{Matthew Grote}
\ead{mgrote@gradcenter.cuny.edu}

\author[add2,add3,add4]{Alexey Ovchinnikov}
\ead{aovchinnikov@qc.cuny.edu}

\author[add5]{Sonia L. Rueda}
\ead{sonialuisa.rueda@upm.es}

\address[add2]{Ph.D. Program in Mathematics, CUNY Graduate Center, New York, USA}
\address[add3]{Ph.D. Program in Computer Science, CUNY Graduate Center, New York, USA}

\address[add4]{Department of Mathematics, CUNY Queens College, Queens, USA}

\address[add5]{Departamento de Matem\'atica Aplicada,
E.T.S. Arquitectura,
Universidad Polit\'ecnica de Madrid, Spain}

\address[add6]{School of Mathematics and Statistics, University of Melbourne, Victoria, Australia}

\begin{abstract}
 Dynamic models describe phenomena across scientific disciplines, yet to make these models useful in application the unknown parameter values of the models must be determined. Discrete-time dynamic models are widely used to model biological processes, but it is often difficult to determine these parameters. In this paper, we propose a symbolic-numeric approach for parameter estimation in discrete-time models that involve  univariate non-algebraic (locally) analytic functions such as $\exp$. We illustrate the performance (precision) of our approach by applying our approach to two archetypal discrete-time models in biology (the flour beetle `LPA' model and discrete Lotka-Volterra competition model). Unlike optimization-based methods, our algorithm guarantees to 
find all solutions of the parameter values  up to a specified precision given time-series data for the measured variables  provided that there are finitely many parameter values that fit the data and that the used polynomial system solver can find all roots of the associated polynomial system with interval coefficients.
\end{abstract}

\begin{keyword}
parameter estimation\sep discrete-time models \sep symbolic-numeric computing \sep  flour beetle model

\MSC[2020] 
92B05\sep
 68W30 \sep 14Q20 \sep 39A60 \sep 13P15

\end{keyword}

\end{frontmatter}

\section{Introduction}
Consider a discrete-time 
model
\begin{equation*}
  \Sigma:%
  \begin{cases}
    {\bm{x}_{t+1}} &= \bm{f}(\bm{x}_t,\bm{\mu},\bm{u}_t),\\
    \bm{y}_t &= \bm{g}(\bm{x}_t,\bm{\mu},\bm{u}_t),\\
    \bm{x}_0 \hspace{-0.3cm} &= \bm{x}^{\ast},
  \end{cases}
\end{equation*}
wherein
\begin{itemize}
\item the state variables $\x$ (a vector) and output variables $\y$
\item  $\f,\g$ are  vector-valued  analytic functions, 
\item input variables $\uu$ (representing e.g., control inputs, etc.),
\item and scalar parameters ${\bm{\mu}},\x^*$, the latter being the initial condition of the system.
\end{itemize}
Discrete-time systems are widely used to model biological processes.
Such systems typically involve unknown parameters.
One is often interested in the values of these parameters due to their biological significance. They may represent key mechanisms such as organism birth rates or survival rates \cite{CDCD}.

\emph{Parameter estimation} is the problem of determining unknown parameters of a model from measured data. In this paper, we propose an algorithm to solve this problem in the case of exact data (data without noise or measurement errors). 
The standard way to find the model parameters is to fit the data with minimal error using an optimization solver over many iterations.
This approach may fail for several reasons: 
e.g., the optimization solver gets stuck at a local minimum, or the solver does find a global minimum but fails to find all global minima in  the case  in which multiple parameter values fit the data exactly. This  
issue can plague 
biological modelers~\citep{COMBOS}.  Our approach is different. It is based on difference algebra, Taylor polynomials to approximate analytic functions, and multivariate polynomial system solving.

There is a partial algorithm~\citep{difference_automatica} to check if multiple parameter values fit the data for discrete-time systems. This approach mimics algorithms for Ordinary Differential Equations (ODEs) based on differential elimination. However, it does not apply to  the large collection of biological models that involve non-algebraic functions (such as the exponential function, integrals, etc.). Such non-algebraic functions exist in biological models ranging from cellular processes to ecological models~\citep{kim2012modeling, marculis2016modelling, adimy2015age}, 
and, in some cases, are mixed with other modeling paradigms such as ODEs and agent-based models (ABMs)~\citep{oremland2015optimal,kim2012modeling}.

Parameter estimation has been a key component in studying flour beetle dynamics~\citep{costantino2005nonlinear,medeiros2014parameter,robertson2009modeling}. The flour beetle system also provides a commonly used test case for a wide range of stage- and age-structured difference equations used in ecology~\citep{beninca2015species, brauer2012mathematical,sullivan2017density}. It is a highly convenient model organism for mathematical modeling, allowing complex dynamics such as chaos that are often predicted in ecological models to 
be observed in data~\citep{brauer2012mathematical,costantino2005nonlinear,robertson2009modeling}. We use a simplified flour beetle model\footnote{This is simplified in that it does not include input variables. This excludes, for instance, observation noise, and also assumes an absence of environmental factors such as predators or food scarcity}~\cite{kuang1996global,cushing2004some, cushing2003cycle,elaydi2010population,cushing2002chaos,robertson2011spatial} to demonstrate our methods. The life cycle of the flour beetle consists of distinct states $L,P,A$, for the larval, pupate, and adult stages respectively. For this reason, we refer to this system as the `LPA model'. Explicitly, this is a discrete-time model wherein by $L_k$ we mean the total larval population at $k$-th observation:
\begin{equation}\label{eq:FB}
 B:\begin{cases}
L_{t+1}=b\cdot A_t\cdot\exp(-c_{EL}\cdot L_t-c_{EA}\cdot A_t)\\ 
P_{t+1}=L_{t}\cdot(1-\mu_{L})\\ 
A_{t+1}=P_t\cdot\exp(-c_{PA}\cdot A_t)+A_t\cdot(1-\mu_A)
\end{cases}
\end{equation}
The parameter $b$ represents the larva birthrate from the surviving eggs, which is proportional to the number of adults. The parameters $\mu_L,\mu_A$ represent the mortality rate of larvae and adults, respectively. The parameters $c_{EL},c_{EA},c_{PA}$ are the rates of cannibalism of eggs by larvae and by adults, and of pupae by adults, respectively. All parameters are nonnegative. We will refer to this system of equations as $B$ throughout the paper.

 We will also consider a series of widely-used competition models from mathematical ecology~\cite{comp_model,cushing2004some}:
\begin{equation}\label{eq:DLVM}
x_i(t+1) = x_i(t)\exp\left(r_i - \sum\limits_{j=1}^na_{ij}x_j(t)\right),\quad i=1,\ldots,n,
\end{equation}
also known as discrete Lotka-Volterra models, where $x_i(t)$ is the $i$th population at time $t$ and $r_i$ and $a_{ij}$ are constant parameters of the model.

The task of this paper is to propose a general algorithm for parameter estimation of a large class of difference equation models. In a system in which $\mathbf{f}$ and $\mathbf{g}$ are analytic, our algorithm approximates the non-polynomial functions by sufficiently close polynomials and then recovers the parameters by symbolic-numeric solving of the resulting polynomial system. There are multiple subtleties in this approach:
\begin{enumerate}
\item The degree of polynomial approximation needs to correspond to the user-specified precision of  parameter estimation,
\item The approximating square polynomial systems are not always consistent, 
\item The resulting consistent square non-linear polynomial systems often have multiple solutions.
\end{enumerate}
In the present paper, we address these 
subtleties
provided that there is an algorithm for polynomial system root isolation in which the input polynomial system has some of the coefficients represented as intervals. As of now, we are only aware of such algorithms as being under development, with the zero problem being one of the bottlenecks.  
We address the second subtlety by 
 checking if the Jacobian of the system is of full rank
and prove the correctness of this criterion for the LPA model  and other models. We address the multiple-solution issue by requiring that the user specifies intervals for the parameter values, and use these intervals to filter the multiple-solution set. (The intervals are also used to assist the Taylor approximation of $\mathbf{f}$ and $\mathbf{g}$, see Section~\ref{sect:experiment} for details.)  Our algorithm can still theoretically return multiple solutions even if all parameters are globally identifiable. However, this has not happened in any of our experiments using real-life biological models. There is a potential to extend this method using over-determined polynomial system solvers to return only one solution in this case, which we leave for future research.
The algorithm is repeatedly demonstrated on the LPA model. We successfully establish a sufficiently small approximation error through numerical experiments.

The paper is organized as follows. In Section~\ref{sect:background}, we review formal concepts in difference equations and parameter estimation, including demonstrative simple models. In Section~\ref{sect:ourapproach},  we {give our precise problem statement and} describe our main algorithm{, whose steps we further describe in subsequent sections. In particular, we present sub-algorithms}
for approximating difference equation systems with analytic functions by polynomial systems in Section~\ref{sec:Taylor}, illustrating this with the LPA model as a running example.  We then discuss errors in our polynomial approximations in Section~\ref{sec:errorTaylor}. In Section~\ref{sec:prolongForSquare}, we show how to create a consistent square system of polynomial equations to recover the parameters  values,
again demonstrating with LPA  and other models. Finally, in Section~\ref{sect:experiment}, we present a randomized numerical experiment on LPA  and competition models showing that varying aspects of our algorithms (namely the polynomial degree and parameter domain size) returns estimated parameter values that are very close to the true parameter values.

\section{Background}\label{sect:background}
\subsection{Systems we consider}
Consider a discrete-time (difference) system of the form
\begin{equation}
  \Sigma:%
  \begin{cases}
    {\bm{x}_{t+1}} &= \bm{f}(\bm{x}_t,\bm{\mu},\bm{u}_t),\\
    \bm{y}_t &= \bm{g}(\bm{x}_t,\bm{\mu},\bm{u}_t),\\
    \bm{x}_0 \hspace{-0.3cm} &= \bm{x}^{\ast},
  \end{cases}
  \label{eq:main}%
\end{equation}
where
\begin{itemize}
  \item $\bm{f}$ and $\bm{g}$ are vectors of functions (combinations of polynomial, rational, exponential and other transcendental functions);
  \item $\bm{x} = \bm{x}_t$ is a vector of state variables that describe the state of the model;
  \item $\bm{\theta} := (\bm{\mu}, \bm{x}^\ast)$ are unknown scalar parameters;
  \item $\bm{u} = \bm{u}_t$ is a vector of input sequence representing ``external forces'', assumed to be known;
  \item $\bm{y} = \bm{y}_t$ is a vector of output variables: this is what we actually observe.
\end{itemize}

For every set $\hat{\bm{\theta}} = (\hat{\bm{\mu}}, \bm{x}^\ast)$ of parameter values and every vector of   
input sequence $\hat{\bm{u}}$ such that the right-hand sides of $\Sigma$ are well-defined, there 
is a unique sequence $\bm{y}$,
denoted $Y(\hat{\bm{\theta}}, \hat{\bm{u}})$.

\subsection{Parameter identifiability: problem statement}
We continue with a statement of the identifiability problem, which is a prerequisite for performing parameter estimation.

\begin{definition}
The parameters $\bm{\theta}$ in a system $\Sigma$ are said to be \emph{globally identifiable} if,
for generic $\hat{\bm{\theta}}$ and generic $\hat{\bm{u}}$,
\begin{equation}
Y(\hat{\bm{\theta}}, \hat{\bm{u}}) = Y(\tilde{\bm{\theta}}, \hat{\bm{u}}) \implies\hat{\bm{\theta}} = \tilde{\bm{\theta}}.
\end{equation}
 \end{definition}
 \begin{remark}
In other words, roughly speaking, the parameters $\bm{\theta}$ are globally identifiable if one can almost always uniquely recover the values of the parameters from observations.
\end{remark}

\begin{remark}
There is a weaker notion of \emph{local identifiability}: the parameters are locally identifiable if one can almost always recover the values of the parameters from observations up to a finite number of options.
\end{remark}

\subsection{Parameter estimation: general problem statement}\label{sec:estprobstat}
 The input for the parameter estimation problem consists of:
 \begin{itemize}
 \item Discrete-time system $\Sigma$,
 \item Data $y_0,y_1,\ldots, y_m$ for $\Sigma$ for some positive integer $m$ corresponding to parameter values $\bm{\hat \theta}$, which are unknown,
 \item Complex balls (real intervals) $R_1,\ldots, R_s$, where $s$ is the length of the tuple $\bm{\theta}$, such that $\bm{\hat\theta_i} \in R_1\times\ldots\times R_s$, and
 \item a real number $\epsilon$.
 \end{itemize}
 The output consists of a finite set of tuples of complex (real) numbers $\bm{\tilde{\Theta}}$\ such that
 there exists $\bm{\tilde \theta} \in \bm{\tilde{\Theta}}$
such that, if $\theta_i$ is locally identifiable, then $|\hat\theta_i-\tilde\theta_i|<\epsilon$. 

\subsection{Parameter estimation: toy examples using typical approach of elimination}
We will show a few toy examples illustrating how one can carry out parameter estimation in simple cases by hand using the standard approach of elimination  (see~\cite{LGEL2009,difference_automatica}).
\begin{example}
Consider a toy model, given by
\[
\begin{cases}
x_{t + 1} = x_t + \mu_1,\\
y_t = x_t,\\
x_0 = \mu_2.
\end{cases}
\]
Both parameters $\mu_1$ and $\mu_2$ can be found from the observed $y_t$ using 
$\mu_1 = y_1 - y_0$ and $\mu_2 = y_0$.
\end{example}
\begin{example}
Consider another toy model, given by
\[
\begin{cases}
x_{t + 1} = x_t + \mu_1+\mu_2,\\
y_t = x_t,\\
x_0 = \mu_3.
\end{cases}\]
Neither $\mu_1$ nor  $\mu_2$ can be separately found from the observed $y_t$.
\end{example}

One can approach the parameter estimation problem for discrete-time models mimicking the existing approach to ODE models by performing an elimination of the unknowns. For instance, let us try to calculate the parameter values in this toy model:
\begin{equation}\label{eq:ex3}
\begin{cases}
x_{t + 1} = \mu_1 x_t +\mu_2,\\
y_t = x_t,\\
x_0 = \mu_3.
\end{cases}
\end{equation}
We will do an elimination of variables to check if we can have an equation relating only $y_t, y_{t+1},\ldots$ and $\mu_1$ and another equation relating only $y_t, y_{t+1},\ldots$ and $\mu_2$. For this, consider the first equation in~\eqref{eq:ex3} and its  shift:
\[
\begin{cases}
x_{t+1} = \mu_1 x_t + \mu_2\\
x_{t+2} = \mu_1 x_{t+1} + \mu_2
\end{cases}.
\]
Substituting 
$x_t = y_t, x_{t+1} = y_{t+1}, x_{t+2} = y_{t+2}$ and performing (Gaussian in this case because the system is linear) elimination of variables, we arrive at the solution
\begin{equation}\label{eq:3sol}
\mu_1 = \frac{y_{t+1} - y_{t+2}}{y_t - y_{t+1}},\quad  \mu_2 = \frac{y_t y_{t+2} - y_{t+1}^2}{y_t - y_{t+1}},
\end{equation}
and so we can uniquely determine the values of $\mu_1$ and $\mu_2$ from the data $\{y_0,y_1,y_2\}$.
\begin{remark} 
This approach is not immediately applicable to
\begin{itemize}
\item larger models because of the computational cost to carry out the elimination of unknowns, 
\item models that involve transcendental functions, e.g., $\exp$, such as model~\eqref{eq:FB}, 
\end{itemize}
and so we arrive at the following challenge:
  design an efficient algorithm for parameter estimation of discrete-time models that involve non-algebraic functions.

\end{remark}

\section{General description of our approach}\label{sect:ourapproach}
In this section, we give our problem statement (as the input/output specification of Algorithm~\ref{alg:expansionrange}) and then, again in Algorithm~\ref{alg:expansionrange}, describe our approach.

By $\bm{y}_{t_i}$, we mean values of the discrete time-dependent output variables, which we call `observed data' as this reflects the measurements a scientist would take in a real-world experiment. In LPA, for example, this would comprise the data $\bm{y}_{t_0}=L_{t_0},~P_{t_0},~A_{t_0}$, and so on.

\begin{algorithm}[H]
\caption{Parameter Estimation}
  \label{alg:PE}\begin{algorithmic}[1]
 
\State {\bf Input}:
\begin{enumerate}
    \item[1] Difference equation system $\Sigma$ whose terms are linear combinations of rational functions and univariate locally analytic functions with computable power series coefficients and whose parameters are locally identifiable.
    \item[2] For each parameter $\theta_i$, 
    user-specified acceptable domain, 
    denoted $R_{\theta_i}$, such that the right-hand sides of $\Sigma$ are defined and so are analytic for all $0\leqslant t\leqslant t_r$ when the parameters take any values in the $R_{\theta_i}$'s.
    \item[3] Observed data $\bm{y}_{t_0}, \bm{y}_{t_1},\ldots,\bm{y}_{t_r}$ corresponding to unknown parameter values $\bm{\hat \theta}$.
    \item[4] Desired error bound $\varepsilon$ for the result.
\end{enumerate}
\State {\bf Output}: Finite set of estimates $\Theta^\ast$ such that there exists $\theta^\ast \in \Theta^\ast$  so that, for all $i$, $|\theta_i^\ast - \theta_i| < \varepsilon$ or failure if there is not enough observed data.

\State Use Algorithm~\ref{alg:expansionrange} to determine the domains over which  all univariate transcendental functions in $\Sigma$ are to be considered.
\State Let $E = \varepsilon$
\State Use Theorem~\ref{thm:findnfortaylor} to determine the degree $N$ and expansion points $\bar a$ of the Taylor polynomials so that the approximation of each analytic non-rational function in $\Sigma$ by a Taylor polynomial of degree $N$ centered at a point from $\bar a$ on the domain of interest has error at most $E$.
\label{step:5}
\State For these  $N$ and $\bar a$, use Algorithm~\ref{alg:dynamictaylor} to find an approximating polynomial system $\Sigma'$.
\State Use Algorithm~\ref{alg:squresystem} to solve $\Sigma'$ using an interval polynomial system solver, given that some of the coefficients of $\Sigma'$ have interval-based entries of width $2E$ and determine the sizes of the intervals for the solutions. If Algorithm~\ref{alg:squresystem} returned failure because of insufficient data, return failure. If there is a size greater than $\epsilon$, then set $E = E/2$, double the precision of the solver, and go to Step~\ref{step:5}.\label{step:seven}
\end{algorithmic}
\end{algorithm}

\begin{theorem} \label{thm:termination}
Algorithm~\ref{alg:PE} terminates if there is a polynomial system solver used in Step~\ref{step:seven} of Algorithm~\ref{alg:PE}.
\end{theorem}
\begin{proof}
This follows from all functions involved being analytic on the domains of interest.
\end{proof}

\begin{remark}\label{rem:4}
As far as we know, a guaranteed interval-based polynomial system solver used in Algorithm~\ref{alg:PE} is still under development. One of the subtleties is the zero problem~\cite{zero} for the coefficients in polynomials appearing in the intermediate computation steps of the solver when a part of the data are intervals.
\end{remark}

In other words,
we replace non-polynomial functions with polynomial approximations by a Taylor series.  We determine relevant domains where a power series expansion can be carried out, then compute enough terms from the Taylor polynomial to be sufficiently accurate.  The details are explained in Sections~\ref{sec:Taylor} and~\ref{sec:errorTaylor}.

We then generate a sufficient number of shifts of the polynomial system to have enough equations to solve for the parameters. We take care of the resulting inconsistency and multiple solution subtleties using full-rank Jacobian condition and solution filtering by user-specified intervals, respectively. See Section~\ref{sec:prolongForSquare} for more details.

Each step is demonstrated using a toy example and in the LPA model based on \cite{CDCD}.
The experimental data in \cite{CDCD} is used to confirm our approach, and further testing was done with ``simulated experimentation,'' generated data within the context of \cite{CDCD}.
Given the model (as in~\eqref{eq:FB}) and data (as in \cite{CDCD} or simulated from \cite{CDCD}), we convert $B$ to a new model $B'$ by approximating $\exp(-c_{EL}L_t-c_{EA}A_t)$ and $\exp(-c_{PA}A_t)$ by Taylor polynomials.  The LPA model is mathematically interesting for our method as it requires to skip some of the equations while generating the shifted polynomial system to avoid inconsistency, which our Jacobian criterion catches. We also show how the approach applies to the series of competition models~\eqref{eq:DLVM}.

\section{Replacement of transcendental function by dynamic approximation}\label{sec:Taylor}

Our first step is to replace each occurrence of a transcendental analytic function with Taylor polynomials. We further restrict that such functions have computable power series coefficients. Each analytic transcendental function in some difference equation system $\Sigma$ we denote by $G_j$. These are not the equations themselves, for example, in the LPA model, the equation \[A_{t+1}=P_t\exp(-c_{PA}A_t)+A_t(1-\mu_A),\] we would call $G_2$ just the term $\exp(-c_{PA}A_t)$. We observe that the argument of any $G_2$ is realized as a function of some of the parameters. 
 For instance, the argument of $G_2$,
denoted  by $\tilde{G}_2$, is $-c_{PA}A_t$ and depends on parameter $c_{PA}$, viewing $A_t$ as a known numerical value.

Of course, the argument $\tilde{G}_j$ in general can be a function of multiple parameters. Again from LPA, we have \begin{gather*}
G_1=\exp(-c_{EL}L(t)-c_{EA}A(t))\\ \tilde{G}_1=-c_{EL}L(t)-c_{EA}A(t).
\end{gather*}
We make the substitution $\tilde{G}_j\to\tau_j\in\mathbb{C}$ to convert the problem to an expansion of the transcendental function of just one variable. In this direction, our first task is determining where to expand these power series.

\subsection{Algorithm for Expansion ranges}
\begin{algorithm}[H]
\caption{Expansion ranges}
  \label{alg:expansionrange}\begin{algorithmic}[1]
\State {\bf Input}:
\begin{enumerate}
    \item[1] Difference equation system $\Sigma$ whose terms are linear combinations of multivariate rational functions and univariate locally analytic functions.
    \item[2] For each parameter $\mu_i$ appearing as an argument of a transcendental function, user-specified acceptable domains for each $\mu_i$, denoted $R_{\mu_i}$.
    \item[3] Observed data $\bm{y}_{t_0},\bm{y}_{t_1},\ldots,\bm{y}_{t_r}$
\end{enumerate}
\State {\bf Output}: Domains $R_{G_{j}}$ over which each transcendental function $G_j$ in $\Sigma$ is expanded.
\Procedure{Expansion range}{} By substitution of maximums and minimums (modulus) of \phantom{-----} $R_{\mu_i}$ into arguments of $G_j$, as relevant, determine radii of discs $R_{L_t}$. 
\State Call the argument of $G_j$ as $\tilde{G}_j=:\tau_j$.
\For{$j=1\to r$, $t$ fixed}
\State $\displaystyle R_{G_{j,t}}:=\left\{\min_{R
_{\mu_i^*,i=1,\ldots,s} }|\tilde{G}_j|\le |\tau_j| \le\max_{R
_{\mu_i^*,i=1,\ldots,s} }|\tilde{G}_j|\right\}$
\EndFor 
\EndProcedure
\end{algorithmic}
\end{algorithm}

\begin{remark}The time-measured data and acceptable parameter intervals are necessary to compute $\min_{R
_{\mu_i^*} }|\tilde{G}_j|,\max_{R
_{\mu_i^*} }|\tilde{G}_j|$. More clearly, the argument $\tilde{G}_j$ varies in both the parameters (unknown) and time (known). This is made especially clear in Section~\ref{ex:workedexample}, where this algorithm is applied to the LPA model.
\end{remark}

To compute a Taylor polynomial, we need three pieces of data: 
\begin{itemize}\item the domain for expansion, 
\item the point to center the expansion, and 
\item the number of terms (or polynomial degree $N$). 
\end{itemize}
Algorithm~\ref{alg:expansionrange} has given the domain. In the general setting, we will allow the user to explicitly provide the center and degree - leaving the automatic choice of these based on the user-specified parameter estimation error of $\epsilon$ (see Section~\ref{sec:estprobstat}) for future research.

\begin{algorithm}[H]
\caption{Dynamic Taylor}
\label{alg:dynamictaylor}
\begin{algorithmic}
\State {\bf Input}:\begin{enumerate}
    \item[1] Data input for Algorithm~\ref{alg:expansionrange}, e.g., system $\Sigma$.
    \item[2] Output of Algorithm~\ref{alg:expansionrange}.
    \item[3] Set number of terms, i.e., the highest degree of Taylor polynomial, $N$.
    \item[4] Expansion point $\alpha_j\in R_{G_j}$ for each $j$.
    \item[5] Time measured data $\bm{y}_{t_0},\bm{y}_{t_1},\ldots,\bm{y}_{t_r}$.
\end{enumerate}
\State {\bf Output}: System of polynomial equations $\Sigma'$ with parameters approximately equal to those of $\Sigma$.
\Procedure{$\Sigma$ to power series $\Sigma^*$}{} Suppose there are $K$ transcendental functions in $\Sigma$.
    \For{$j=1\to K$}
    \State $\tau_j:=\tilde{G}_j$
    \State $G_j\to T_{G_j}:=\sum_{k=0}^\infty G_j^{(k)}(\alpha_j)(\tau_j-\alpha_j)^k$, where differentiation is with respect to $\tau_j$.
    \State For every equation $e_i=G_{i_1}+\cdots+G_{i_r}+\Delta_i$ in $\Sigma$, $e_i^*=T_{G_{i_1}}+\cdots+T_{G_{i_r}}+\Delta_i$ is in $\Sigma^*$, \phantom{---------}where $\Delta_i$ is  rational.
    \EndFor
\EndProcedure
\Procedure{$\Sigma^*$ to rational $\Sigma'$}{}
    \For{$j=1\to K$}
    \State $\hat{T}_{G_j}:=\sum_{k=0}^N G_j^{(k)}(\alpha_j)(\tau-\alpha_j)^k$
    \State $\tau_j\mapsto \tilde{G}_j$
    \State For every equation $e_i^*=T_{G_{i_1}}+\cdots+T_{G_{i_r}}+\Delta_i$ in $\Sigma^*$, $e_i'=\hat{T}_{G_{i_1}}(\tilde{G}_{i_1})+\cdots+\hat{T}_{G_{i_r}}(\tilde{G}_{i_r})+\Delta_i$ \phantom{---------}is in $\Sigma'$, where $\Delta_i$ is rational.
    \EndFor 
    \State Substitute $\bm{y}_{t_0},\bm{y}_{t_1},\ldots,\bm{y}_{t_r}$ throughout $\Sigma'$ as they appear, if known.
\EndProcedure
\end{algorithmic}
\end{algorithm} 

\begin{remark}
Note that the algorithm considers every transcendental term in the system separately. For example, the LPA system is written as three equations, but represents an indefinite number of equations all depending on $t$. Hence for instance it includes both the equations \[L_1=bA_0\exp(-c_{EL}L_0-c_{EA}A_0), \quad L_2=bA_1\exp(-c_{EL}L_1-c_{EA}A_1)\] and considers \[G_1=\exp(-c_{EL}L_0-c_{EA}A_0)\] as a replacement in the equation for $L_1$ while \[G_{2}=\exp(-c_{EL}L_1-c_{EA}A_1)\] would be used for $L_2$.
\end{remark}

\begin{remark}
 Note that $G_j$ actually depends on $t$ as well as $\tau$, for instance $G_1$ and $G_2$ above substitute $L_t,A_t$ for times $t=0,1$ respectively. For readability and enumeration in the algorithm, this aspect is suppressed: substitution of time measured data (e.g. values $L_t,A_t$ above) is intrinsic in the algorithm.
\end{remark}

\subsection{Worked example}\label{ex:workedexample}

We demonstrate our method using the LPA model~\eqref{eq:FB}. Note that all numerical values in this section have been rounded to significant figures for readability, and exact values for this example are available on GitHub as a Maple worksheet, \url{https://github.com/jedforrest/larva-pupa-adult/blob/main/LPA_example_Maple.mpl}.

\begin{table}[h]
    \centering
    \begin{tabular}{|c||c|c|c|}\hline
       $  t$&$L_t$&$P_t$&$A_t $ \\\hline
         0& 107& 73& 214\\\hline
         1& 33&86 &240 \\\hline
         2& 67& 27& 267\\\hline
    \end{tabular}
    \caption{Population data for Section~\ref{ex:workedexample}}
    \label{table:workedexample}
\end{table}
Suppose that we have observed the population data in Table~\ref{table:workedexample}, and $\Sigma$ is the LPA model~\eqref{eq:FB}.
\begin{remark}
 If data are available for more values of $t$ than, e.g., in Table~\ref{table:workedexample},  this data may over-determine the system and cause it to be inconsistent. Therefore,  to stay within consistent square systems of polynomial equations  our approach will disregard this extra data. We leave it for the future research to incorporate over-determined polynomial system solvers into our approach to take advantage of additional data.
\end{remark}
To apply Algorithm~\ref{alg:dynamictaylor}, we first 
 use Algorithm~\ref{alg:expansionrange}. 
Note in this model that all parameters are positive real numbers, hence the domains $R_{\mu_i}$ become intervals. Suppose the user has determined acceptable ranges:
\begin{equation}
\label{tab:acceptablerangesexample}
\begin{aligned}
    c_{EL}&\in [0.01,0.014]=R_{c_{EL}}, \\
    c_{EA}&\in[0.0097,0.0134]=R_{c_{EA}},\\
    c_{PA}&\in[0.0032,0.0062]=R_{c_{PA}}
\end{aligned}
\end{equation}
We determine expansion ranges for the $\exp$ term in $L_t$ by finding its endpoints with the formulas:
\begin{equation}
\label{eq:minmax}
\begin{aligned}
    \min R_{L_{t}}&=-\max R_{c_{EL}}L_t-\max R_{c_{EA}}A_t,\\\max R_{L_{t}}&=-\min R_{c_{EL}}L_t-\min R_{c_{EA}}A_t.
\end{aligned}
\end{equation}
 Substituting the data from Table~\ref{table:workedexample} and~\eqref{tab:acceptablerangesexample} into~\eqref{eq:minmax}, we obtain
\begin{gather*}\max R_{L_1}=-0.010\cdot 107-0.01\cdot 214,\\
\min R_{L_1}=-0.014\cdot 107-0.013\cdot 214,
\end{gather*}
hence \[R_{L_1}=[-3.67,-2.67].\] Similarly $R_{A_t}$ is found via the formulas:
\begin{align*}
    \min R_{A_{t}}&=-\max R_{c_{PA}}A_t,\\\max R_{A_t}&=-\min R_{c_{PA}}A_t.
\end{align*}

No other user input is required for Algorithm~\ref{alg:expansionrange} because no other parameters appear in transcendental functions. Hence they are not necessary for Algorithm~\ref{alg:dynamictaylor}, however, they may be of use in the future Algorithm~\ref{alg:squresystem}.

The results of Algorithm~\ref{alg:expansionrange} are recorded in Table~\ref{table:exampleranges}, as well as the midpoints $\alpha_t,\beta_t$ respectively. These are our expansion points ($\beta$ is used instead of repeating $\alpha$ for readability), and the use of midpoints is justified briefly in the next sections.

\begin{table}[h]
    \centering
    \begin{tabular}{|c||c|c|c|c|}\hline
        t &$R_{L_t}$&$\alpha_t$&$R_{A_t}$&$\beta_t$  \\\hline
         1&$[-3.67,-2.67]$& $-3.17$&$[-1.48,-0.77]$& $-1.13$\\ \hline
         2&$[-4.50,-3.28]$& $-3.89$&$[-1.65,-0.86] $&$-1.25$\\\hline
    \end{tabular}
    \caption{Expansion ranges and midpoints}
    \label{table:exampleranges}
\end{table}
We move on to apply Algorithm~\ref{alg:dynamictaylor}. For readability, we choose $N=2$ as the degree of the expansion. Note the accuracy rises with increasing degree, as we find in Section~\ref{sect:experiment}. For illustration, every step is carried out for $t=1$ below. Note $\tau_1=-c_{EL}L(t)-c_{EA}A(t)$, $\tau_2=-c_{PA}A(t)$.
\begin{align*}
T_{L_1}&=\sum_{k=0}^\infty\frac{1}{k!}(\tau_1-(-3.17))^k=0.04 + 0.04(\tau_1+ 3.17) + 0.02(\tau_1+ 3.17)^2+O(\tau_1^3)\\T_{A_1}&=\sum_{k=0}^\infty\frac{1}{k!}(\tau_2-(-1.13))^k=0.32 + 0.32(\tau_2 + 1.13) + 0.16(\tau_2 + 1.13)^2 + O(\tau_2^3)
\end{align*}
By choice of $N=2$, this yields \begin{gather*}\hat{T}_{L_1}=0.04 + 0.04(\tau_1+ 3.17) + 0.02(\tau_1+ 3.17)^2 \\\hat{T}_{A_1}=0.32 + 0.32(\tau_2 + 1.13) + 0.16(\tau_2 + 1.13)^2
\end{gather*}
Repeating this process for $t=2$ yields the data of 
Table~\ref{table:examplepolynosubs}.
\begin{table}[h]
    \centering
    \begin{tabular}{|c||c|c||c|}\hline
        t & $\hat{T}_{L_t}$ & $\hat{T}_{A_t}$\\\hline
        1 & $0.04 + 0.04(\tau_1+ 3.17) + 0.02(\tau_1+ 3.17)^2$ & $0.32 + 0.32(\tau_2 + 1.13) + 0.16(\tau_2 + 1.13)^2$\\\hline
        2 &$0.02 + 0.02(\tau_1+3.89) + 0.01(\tau_1 + 3.89)^2$ &$0.29 + 0.29(\tau_2+1.25) + 0.14(\tau_2 + 1.25)^2$ \\\hline
    \end{tabular}
    \caption{Taylor polynomials in $\tau_1,\tau_2$}
    \label{table:examplepolynosubs}
\end{table}
Finally, we reverse the substitutions \[\tau_1\mapsto-(c_{EL}L_{t}+c_{EA}A_{t}),\quad\tau_2\mapsto-c_{PA}A_{t}\] to create our full, multivariate polynomial system~\eqref{eq:examplesystem}, denoted  by $B'$.  Since $L_t, A_t$ 
 will be substituted with data,
$\tau_1,\tau_2$  can both vary on each time step.

For example, the first three equations of system $B'$ are
the initial conditions $L_0=107,~ P_0=73,~A_0=214$. To find $L_1$, we note  that \begin{gather*}\tau_1\mapsto-(c_{EL}L_0+c_{EA}A_{0})=-107c_{EL}-214c_{EA},\\ L_1=33\end{gather*} and hence in $B'$:
\begin{align*}
    33&=214b\left(0.18 + 0.04(-107c_{EL}-214c_{EA})+ 0.02(-107c_{EL}-214c_{EA}+ 3.17)^2\right)
\end{align*}
Similarly $\tau_2\mapsto -c_{PA}A_0=-214c_{PA}$, $A_0=214$, $P_0=73$ and $A_1=239$ yields:
\begin{align*}
    239=73\left(0.69 + 0.32(-214c_{PA})+ 0.16(-214c_{PA} + 1.13)^2\right)+214\cdot(1-\mu_A)
\end{align*}
No expansion occurs in $P_t$ as it has no transcendental functions. However, data substitutions are carried out. Continuing this process completes Algorithm~\ref{alg:dynamictaylor}. The result is below:
\begin{equation}\label{eq:examplesystem}
    B':\begin{cases}
        L_0=107\\
        P_0=73\\
        A_0=214\\
L_1=33=214b\left(0.18 + 0.042(-107c_{EL}-214c_{EA})\right. \\ \qquad\qquad+ \left.0.02(-107c_{EL}-214c_{EA}+ 3.17)^2\right)\\
        P_1=86=107(1-\mu_{L})\\
        A_1=239=73\left(0.69 + 0.32(-214c_{PA})\right. \\ \left. \qquad\qquad+ 0.16(-214c_{PA} + 1.13)^2\right)+214\cdot(1-\mu_A)
        \\
        L_2=67 = 240b(0.1 - 0.67c_{EL} - 4.89c_{EA} + 0.01(-33c_{EL} - 240c_{EA} + 3.89)^2)\\
        P_2=27= 33(1-\mu_{L})\\
        A_2=267 = 55.29 - 5884.55c_{PA} + 12.26(-240c_{PA} + 1.25)^2 + 240(1-\mu_A)
    \end{cases}
\end{equation}
This comprises 9 equations. However,  the first three provide no new information and can be
 ignored
when solving the system for parameters. 
Note that the equations for $P_1, P_2$ cause the system to be inconsistent, as the former solves $\mu_L=22/107$ while the latter yields $\mu_L=7/33$, which are not equal. This issue is addressed in  Algorithm~\ref{alg:squresystem}, and this example continues in Section~\ref{ex:continuedexample}.

\section{Error of Taylor approximation}\label{sec:errorTaylor}

In Algorithm~\ref{alg:dynamictaylor}, the user specifies the number of terms for the Taylor polynomial. However, as mentioned  above, this number may also be determined by use of the Taylor remainder. This is discussed in this section. Here the user is providing an acceptable error in the gap of values between an equation in given difference equation system $\Sigma$ and its approximation formed by replacing transcendental terms by Taylor polynomials.
\begin{example}
Suppose that we want to expand the term $\exp(-c_{PA}A_2)$ in the equation for $A_3$ in some system. If the user has provided that $c_{PA}\in[0.01,0.027]$ (in other words, $R_{c_{PA}}=[0.01,0.027])$ and $A_2=10$, then Algorithm~\ref{alg:expansionrange} finds that $\tau=c_{PA}A_2$ lies in the interval $R_{A_3}=[0.1,0.27]$. As in Algorithm~\ref{alg:dynamictaylor}, we shall use the midpoint of this interval, $a=0.185$, as the point of expansion.
If the user specifies the acceptable error of $\varepsilon=0.01$,~\eqref{eq:errorineq} yields:
\begin{equation*}
    \frac{e^{-0.1}(0.185)^{N+1}}{(N+1)!}<0.01,
\end{equation*}
which translates into $N > 1.16547$, and so $N=2$  guarantees that the Taylor series has an error less than 0.01 in this system.
\end{example}

Because we approximate transcendental terms by (Taylor) polynomials, by substituting these into the original equation the new (approximating) equation may in general be rational. This will depend on the form of the original equation from $\Sigma$ which is to be approximated.

\subsection{Analysis in LPA}

As a start, we demonstrate the use of Taylor remainder in determining the requisite number of terms for Taylor polynomials with the LPA model first. Note that the $\exp$ in the equations of both $L_t$ and $A_t$ receives a negative number as input, hence we analyze the function $e^{-\tau}$.
Suppose that $e^{-\tau}$ is expanded over interval $[\alpha,\beta]$ at point $a\in[\alpha,\beta]$ with given error $\varepsilon>0$. Then for any $x\in[\alpha,\beta]$, there is a value $\xi$ between $x,a$ such that:
\begin{align*}
    \left|(-1)^{N+1}e^{-\xi}\frac{(x-a)^{N+1}}{(N+1)!}\right|&<\varepsilon
\end{align*}
which simplifies to:
\begin{equation*}
    \frac{e^{-\xi}|x-a|^{N+1}}{(N+1)!}<\varepsilon
\end{equation*}
This value $\xi$ maximizes the value of $e^{-\tau}$. Since $e^{-\tau}$ is a monotonically decreasing function,
$\xi=\min\{x,a\}$. Furthermore $e^{-\alpha}>e^{-\tau}$ for any $\tau\in(\alpha,\beta]$ so that we can instead solve:
\begin{equation*}
    \frac{e^{-\alpha}|x-a|^{N+1}}{(N+1)!}<\varepsilon
\end{equation*}
We need to handle the maximum possible error, so we next consider the maximum value of $|x-a|^{N+1}$. View $|x-a|$ as the length of the interval between $x,a$. Hence as $x$ ranges over $[\alpha,\beta]$, the maximal value of $|x-a|$ is either $|\alpha-a|$ or $|\beta-a|$.
Combining these, our inequality in the Taylor remainder has only $N$ as a variable:
\begin{equation}\label{eq:errorineq}
    \frac{e^{-\alpha}\max\{|\alpha-a|,|\beta-a|\}^{N+1}}{(N+1)!}<\varepsilon
\end{equation}
Thus, if the user instead provides an acceptable level of error $\varepsilon$, the appropriate number of terms (or degree of Taylor polynomial) can be determined. This is not currently implemented in our algorithm as our numerical experiments involved observing the effects of the Taylor degree (See Section~\ref{sect:experiment}); however, it is straightforward to include this calculation in the future.

\subsection{Choice of midpoint for expansion}

A reader exploring our numerical experiments in Section~\ref{sect:experiment} will note our choice to use the midpoint of various intervals as the point of Taylor polynomial expansion. Both experiments investigate systems involving $\exp$ and hence we justify our choice in this section.

In Algorithm~\ref{alg:dynamictaylor}, the Taylor series are all expanded at the midpoints of $R_{L_t},R_{A_t}$, though other expansion points are possible. Our rationale for choosing the midpoints follows from minimizing:
\begin{align*}
    \min_{a\in[\alpha,\beta]}\frac{e^{-\alpha}\max\{|\alpha-a|,|\beta-a|\}^{N+1}}{(N+1)!}
\end{align*}
where all notation follows from the previous section. In this problem, only $a$ may vary. Since $\alpha\le a\le \beta$, it is straightforward to see that $\max\{|\alpha-a|,|\beta-a|\}$ is minimized when $a=\dfrac{\alpha+\beta}{2}$, that is, when $a$ is halfway between either end of the interval.

\subsection{Generalizing the replacement of transcendental functions}\label{sec:Taylor-gen}

Suppose that $e$ is some equation in a difference equation system $\Sigma$ which involves analytic transcendental functions whose arguments are multivariate polynomials in the states, inputs, and/or parameters. 

\subsubsection{Single transcendental function in an equation}
Consider first that $e$ has exactly one such transcendental function $f$ as a term and call its argument $\tau$. 

\begin{example}

In LPA, the equation $A_{t+1}=P_t\cdot\exp(-c_{PA}\cdot A_t)+A_t(1-\mu_A)$ can play the role of $e$, with $f(\tau)=\exp(-c_{PA}\cdot A_t)$ and $\tau=-c_{PA}\cdot A_t$.

\end{example}

\begin{lemma}\label{lem:maxexists}
    $|f^{(N+1)}(\tau)|$ achieves its maximum on the intervals established by algorithm ~\ref{alg:expansionrange}.
\end{lemma}

\begin{proof}
    Note that $\tau:\mathbb{R}^r\to\mathbb{R}$ is a continuous function (better, polynomial). Now, our algorithm $\hyperlink{Expansion ranges}{\ref{alg:expansionrange}}$ provides valid intervals for every term in any possible $\tau$, by construction. Calling the product of these $I$, note then that $I$ is a closed, bounded subspace of $\mathbb{R}^r$ hence compact, and $\tau$ is continuous. Thus $\tau(I)$ is compact.

    Note $f$ is analytic and hence $f^{(N+1)}$ is continuous. Hence $f^{(N+1)}$ can be viewed as a continuous function $\mathbb{R}\to\mathbb{R}$ over the compact set $\tau(I)$. The continuity of $f^{(N+1)}(\tau)$ implies the continuity of $|f^{(N+1)}(\tau)|$ and hence over compact $\tau(I)$ this last function achieves its maximum.
\end{proof}

Call the value of $\tau$ maximizing $|f^{(N+1)}(\tau)|$ by $\xi$. Suppressing the data of $\tau$ as in the proof of Lemma~\ref{lem:maxexists}, we view $f(\tau)$ not as $\mathbb{R}^r\to\mathbb{R}$, but as a function $f:\mathbb{R}\to\mathbb{R}$. The error then in replacing $f$ by its power series in $\tau$ follows from the Taylor remainder formula:
\begin{align*}
    R_N(\tau)&=\frac{f^{(N+1)}(\xi)}{(N+1)!}~(\tau-a)^{N+1}
\end{align*}
where $a$ is an expansion point in $\tau(I)$. The task of minimizing this error $R_N$ for a given $\epsilon>0$ becomes:
    \begin{align*}
       \left| \frac{f^{(N+1)}(\xi)}{(N+1)!}(\tau-a)^{N+1}\right|&<\epsilon.
    \end{align*}
 
\begin{theorem}\label{thm:findnfortaylor}
    Given $\epsilon>0$, $\xi$ as above, the maximum value of $\tau$ over $I$, expansion point $a\in\tau(I)$, and $|f^{(N+1)}(\xi)|$, there is a computable value $N$ so that the $N$-th Taylor polynomial of $f(\tau)$ is within $\epsilon$ error of the exact value.
\end{theorem}

\begin{proof}

    Note $\xi$ exists by Lemma~\ref{lem:maxexists}. Assuming $|f^{(N+1)}(\xi)|$ is known, our task in finding $N$ becomes solving:
    \begin{align*}
        \max_{I}\left|\frac{f^{(N+1)}(\xi)}{(N+1)!}(\tau-a)^{N+1}\right|&<\epsilon
    \end{align*}
    for the minimum $N$ making the inequality true.
    
    By the triangle inequality and definition of $\xi$, we achieve:
    \begin{align*}
        \max_{I}\left|\frac{f^{(N+1)}(\xi)}{(N+1)!}(\tau-a)^{N+1}\right|&\le\frac{|f^{(N+1}(\xi)|}{(N+1)!}\left(\max_I\left|\tau\right|+|a|\right)^{N+1}
    \end{align*}
    and hence instead solve:
    \begin{align*}
        \frac{|f^{(N+1}(\xi)|}{(N+1)!}\left(\max_I\left|\tau\right|+|a|\right)^{N+1}&<\epsilon
    \end{align*}
    Because the values $|f^{(N+1)}(\xi)|,\max_I\left|\tau\right|,|a|$ are determined (and are real numbers), the left hand side is now a function of $N$ only. The inequality can be solved for instance by treating $N$ as real valued, solving:
    
    \begin{align}\label{eq:solveforn}
        \frac{|f^{(N+1)}(\xi)|}{(N+1)N(N-1)\cdots1}\left(\max_I\left|\tau\right|+|a|\right)^{N+1}&=\epsilon
    \end{align}
    and calling $y$ the minimum of real positive solutions, setting $N=\lceil y\rceil$.
    \end{proof}

    \begin{remark}\label{rem:aboutfindforntaylor}
        Note~\ref{thm:findnfortaylor} makes big asks. Some are easier: our ability to maximize continuous $|\tau|$ over a compact domain is not difficult. In our setting $\tau$ is a multivariate polynomial over a product of closed intervals, and this task is fairly easy.

        If we make the assumption that $f$ is a computable power series, we can access $f^{(N+1)}(\tau_0)$ for any fixed value $\tau_0$ in the domain of $f$. Consequently since $\xi$ exists by ~\ref{lem:maxexists} we can access $|f^{(N+1)}(\xi)|$.
        
        To find $\xi$, suppose further that access to $f^{(N+1)}(\tau_0)$ gives us a formula for $f^{(N+1)}(\tau)$. Letting this take values on $\tau(I)$, we simply find $\xi\in\tau(I)$ which maximizes $|f^{(N+1)}(\tau)|$, i.e., again maximizing a continuous function over a compact domain.

        In practical application these assumptions are met frequently. Elementary transcendental functions such as $\sin,~\cos,~\log$ or $\exp$ have computable power series, formulas for coefficients, and are easy to maximize over compact domains.
    \end{remark}

\begin{example}
    
 Continuing in the demonstration via the LPA model, suppose we wish to find the number of expansion terms in algebraically approximating:
    \begin{align*}
        L_{2}=b A_t\exp(-c_{EL}L_1-c_{EA} A_1)
    \end{align*}
    The function $\tau=-c_{EL}L_1-c_{EA} A_1$ and the function $f=\exp$. Then $\tau$ is a multivariate polynomial, hence continuous everywhere and achieving its maximum on any closed domain. Further $\exp$ is analytic everywhere, satisfying our requirements. We find that $|f^{(N+1)}(\tau)|=\exp(\tau)$ in this example.

    For illustration purposes suppose the user inputs/finds $I_{\mu}=[0.1,0.5]$ for all parameters $c_{EL},c_{EA},L_1,A_1$. Then $\max_I\tau=-0.02$. Because $\exp$ is monotonic it follows that $\xi=-0.02$ as well. Suppose the user expands at $a=0.4$ in $[0.1,0.5]$ and will tolerate an error of $\epsilon=0.01$. Then we must find the minimal $N$ so that:
    \begin{align*}
        \frac{\exp(-0.02)}{(N+1)!}(0.42)^{N+1}<0.01
    \end{align*}
    from which we determine $N=3$, so only 3 terms are needed in the expansion.

    \end{example}

    \begin{example}
        
Suppose we encounter the equation:
    \begin{align*}
        E_{t+1}&=AE_t+\log(bE_{t-1})
    \end{align*}
    and wish to approximate in our method. For illustration, suppose $I_\mu=[0.02,0.03]$ for all parameters $A,b,E_{t-1}$.
    
We must find $\xi$ maximizing the absolute value of the $N+1$-th derivative of $\log$, which is $|\tau|^{-N-1}N!$. For fixed $N$, when $0<|\tau|<1$, this is maximized when $|\tau|$ is minimized. Observe that $\min_I\tau=0.004$, hence, $\xi=0.004$.
    
    Note also $\max_I\tau=0.009$, and $\tau$ takes all values between; so $\tau(I)=[0.004,0.009]$. Taking $a=0.005$, supposing $\epsilon=0.002$,
    \begin{align*}
            \frac{|0.004|^{-N-1}/N!}{(N+1)!}(0.009)^{N+1}&<0.002
    \end{align*}
    from which we find $N=5$.
     \end{example}

\subsubsection{Multiple transcendental functions in an equation}
    Let us now suppose $e=\Delta+f_1(\tau_1)+\cdots+f_k(\tau_k)$ where $\Delta$ is rational, $f_i$ are transcendental and $\tau_i$ are rational functions depending on the states, inputs, parameters.

    Calling $e^*=A+T_1+\cdots+T_k$  the algebraic approximation (by rational functions) of $e$ (hence $T_i$ are Taylor series to $f_i$) our task in broadest terms is to ensure:
    \begin{align*}
        |e-e^*|<\epsilon
    \end{align*}
   for given $\epsilon>0$. Note the rational term $A$ in $e$ are present in $e^*$ so will cancel. By the triangle inequality certainly
   \begin{align*}
       |e-e^*|&\le \sum_{i=1}^k|f_i-T_i|
   \end{align*}
   so that we may minimize the right hand side. In particular if we achieve for all $i$:
   \begin{align*}
       |f_i-T_i|<\frac{\epsilon}{k}
   \end{align*}
   then the result holds. In general, we may relax to $|f_i-T_i|<c_i\epsilon$ satisfying $c_1+\cdots+c_k=1$, $c_i>0$ for all $i$.

\begin{example}
    
Suppose we desire to to approximate this equation by rational functions:
    \begin{align*}
        E_{2}&=AE_t+\exp(bE_1)+\log(dE_1)
    \end{align*} 
    Suppose the user inputs/finds $I_\mu=[-0.1,0.3]$ for $\mu=a,b,E_1$, $I_d=[0.1,0.4]$ and $\epsilon=0.01$. Following notation $f_1=\exp(bE_1),f_2=\log(dE_1),\tau_1=bE_1,\tau_2=dE_1$.
    
    Then $\xi_1=\max\tau_1=0.09$ by monotonicity of $\exp$, which is the absolute value of $N+1$-th derivative of $\exp$. The absolute value of the $N+1$-th derivative of $\log$ is $|\tau|^{-N-1}N!$. For fixed $N>1$, when $0<|\tau|<1$, this is maximized when $|\tau|$ is minimized. Observe that $\min_I\tau=0.01$, hence $\xi_2=0.01$ (arguing again by monotonicity). Note finally that $\max\tau_2=0.16$.

 Suppose we expand at $a_1=0.08, a_2=0.1$.
    Our task is to find $N_1,N_2$:
    \begin{align*}
        \frac{\exp(0.09)}{(N_1+1)!}(0.17)^{N_1+1}&<0.005\\\frac{|0.1|^{-N_2-1}/N_2!}{(N_2+1)!}(0.026)^{N_2+1}&<0.005
    \end{align*}
from which $N_1=1$ and $N_2=2$.
This provides us the polynomial approximation:
\begin{align*}
    E_{2}'&=AE_1+\left(1+(bE_1-0.08)\right)+\left(-2.30259 + 10 (dE_1 - 0.1) - 50 (dE_1 - 0.1)^2\right)
\end{align*}
\end{example}

\subsubsection{Picking expansion point}

In the above generalization we treated expansion point $a$ as a fixed choice. Here we offer some insight for a user interested in picking expansion point $a$.

One approach is to analyze the equation of the Taylor remainder~\eqref{eq:solveforn} as a 2-variable function in $a,N$. Note per Remark~\ref{rem:aboutfindforntaylor} the other terms in~\eqref{eq:solveforn} may be treated as given, leaving the appropriate $a,N$ as the only unknown terms. Now apply available root solving methods, noting $a$ varies over $\tau(I)$ and $N\in\mathbb{N}$. The `solve' command in Maple or `IntervalRootFinding.jl' in Julia provide options, the latter with the issue that a bounded interval must be given for $N$. (This may be reasonable however since $N$ represents a number of terms in a Taylor polynomial, and computing power naturally limits extreme values of $N$.) We did not include this approach in the scope of our project per our choice of midpoint when implementing the algorithms in Julia.

Alternately the user may pre-select $N$ and solve for $a$, analogous to the previous sections where we pre-selected $a$ and solved for $N$. The issue being a solution for $a$ may not exist: it is possible that too few terms (selected $N$) in a Taylor polynomial will never be sufficient for a user provided error $\epsilon$.

\begin{example}
    If $f(\tau)=\log(\tau)$ is expanded over $\tau(I)=[0.1,0.3]$ within error $\epsilon=0.01$ then we would solve:
    \begin{align*}
        \frac{0.1^{-N-1}/N!}{(N+1)!}(0.3+a)^{N+1}&=0.01
    \end{align*}
    (note $|a|=a$ because $a\in[0.1,0.3]$). Analyzing the left hand side by letting $a$ vary and solving for $N$ reveals that $N$ must exceed 5. 

Inverting the process for emphasis, if the user had pre-selected $N=3$ then by substitution the problem becomes:
\begin{align*}
    69.4444 (0.3 + a)^4=0.01
\end{align*}
whose solutions $a$ are either negative or complex, in every case not valid as $a\in[0.1,0.3]$.
\end{example}

In our implementation, and in general, the apparent majority of transcendental functions of interest converge rapidly within relatively small values of $N$, see Section~\ref{sect:experiment}. It is generally safe for the user to pre-select $a$, typically as a midpoint as we have done, and repair at issue holistically.

\subsubsection{Error in parameter values}

This section has concerned itself with error $\epsilon>0$ between an equation $e$ in difference equation system $\Sigma$ approximated by rational functions to $e'$. Of course, the ultimate problem is to solve for parameters from $e'$. What then is the error on the parameters? A deep analysis is beyond the scope of this paper but we address some thoughts as follows:

\begin{itemize}
    \item We will denote the solutions of (parameters found from) $\Sigma'$, which approximates $\Sigma$, by $\hat\mu$. 
    \item The set $S$ of all solutions $\hat\mu$ of $\Sigma'$ is 
    finite. This is because the $\Sigma^\ast$ is selected so that its Jacobian is of full rank, and so the corresponding affine variety is of zero dimension.
    \item The user has provided guess intervals $R_{\mu_k}$ for each $\mu_k$ of ${\bf \mu}$. We `filter' $S$ by intersecting with these intervals to remove impermissible solutions. This is formally part of Algorithm~\ref{alg:squresystem}.
    \item If $e'$ is the equation in $\Sigma'$ approximating $e$ from $\Sigma$, note $e'(\hat\mu)=0$ by definition. Then the error $|e(\hat\mu)-e'(\hat\mu)|<\epsilon$ tells us that $e({\hat\mu})\in B_\epsilon$, the ball center 0 radius $\epsilon$.
    \item Call $\tilde{\mu}$ the actual parameter solution. By construction $\tilde{\mu}\in e^{-1}(B_\epsilon)$. Intersecting $e^{-1}(B_\epsilon)$ with all of the $R_{\mu_k}$ yields the space where both $\tilde{\mu},\hat\mu$ both live. Call this set $D$.
    \item Hence, the maximal error in our algorithm is the maximum over $e$ in $\Sigma$ of $\diam(D)$.
    \item The error to any particular $\widetilde{\mu_k}$ is found by projecting onto $\mu_k$ in parameter space, that is, maximum across $e$ in $\Sigma$ of $\diam(\proj_{\mu_k}(D))$.
    \item Describing $D$ therefore involves analysis on the fibers of $e$ in $\Sigma$, which is beyond our scope.
    \item The user has authority in limiting $\diam(D)$ by choice of $R_{\mu_k}$. This fits the intuition that `a better guess yields better results.' In Section~\ref{sect:experiment}, we analyze performance of the algorithm with increasing $R_{\mu_k}$ in some real world biological models. Indeed, broader guess intervals $R_{\mu_k}$ increase `difficulty' e.g., required higher degrees of Taylor polynomial for convergence.
    \item However, the relatively good approximation of Taylor polynomial terms pushes $\hat\mu$ to $\tilde{\mu}$ even in relatively low degree $N$ in many cases. See the numerical experiments in Section~\ref{sect:experiment}, which demonstrate convergence to a large number of significant digits quite rapidly.
    \item There is also error encountered via the specific method of solving the multivariate polynomial system $\Sigma^*$. This is left as a blackbox in Algorithm~\ref{alg:squresystem}, and methods by approximation could invite more error. However, filtration through the $R_{\mu_k}$ reveals the set of interest to be exactly the set $D$ already discussed.
\end{itemize}

\section{Shifts to obtain a consistent square polynomial system}
\label{sec:prolongForSquare}
\subsection{Theoretical guarantee for consistency: explicit condition  beyond full-rank Jacobian}
We saw in Section~\ref{ex:workedexample} that the shift process can create inconsistent square polynomial systems.
In this section, we demonstrate a technique of shift of a difference equation system to guarantee a solution.

The systems we will consider from now on are of the form:
\begin{equation}
  \Sigma:%
  \begin{cases}
    {\bm{x}_{t+1}} = \bm{f}(\bm{x}_t,\bm{\mu}),\\
  \y_t = \x_t\end{cases}\label{eq:main_simplified}%
\end{equation}
Comparing with the general system in \eqref{eq:main},  we assume that there are 
no input variables $\uu_t$ and that the values of all state variables are known.

Consider the output system $\Sigma'$ of Algorithm~\ref{alg:dynamictaylor}. This is a system of polynomial equations 
\begin{equation}\label{eq:Sigmap}
  \Sigma':  \bm{F} = {\bm{x}_{t+1}}- \bm{p}(\bm{x}_t,\bm{\mu})
\end{equation}
with a vector $\bm{x}=(x_1,\ldots ,x_n)$ of state variables all substituted with known values,  a vector $\bm{\mu}=(\mu_1,\ldots ,\mu_{L})$ of unknown scalar parameters, and a vector $\bm{p}=(p_1,\ldots ,p_n)$ of polynomials in $\mathbb{C}[\bm{\mu}] =\mathbb{C}[\mu_1,\ldots ,\mu_{L}]$. 
Let us call $\sigma$ the automorphism that takes $\bm{x}_t$ to $\bm{x}_{t+1}$ and study how to use it to extend the system  $\Sigma'$  to define a system $\sigma\Sigma'$ of $L$ polynomial equations in the $L$ scalar parameters $\bm{\mu}$. More explicitly, we write $\bm{x}_t$ as $(x_1(t),\ldots ,x_n(t))$ and the system $\Sigma'$ as
\[
  \Sigma':%
  \begin{cases}
   F_1 &= {x_{1}(t+1)}- p_1(x_1(t),\bm{\mu}),\\
    &\vdots \\
   F_n  &= {x_{n}}(t+1)-p_n(x_n(t),\bm{\mu}).
  \end{cases}
  \qedhere
\]
To shift the system, formally we would have to consider two steps:
\begin{itemize}
\item First assume that $x_j$, $j=1,\ldots ,n$ are generic difference variables to apply the automorphism $\sigma$, denote $x_j$ as $x_{j,0}=x_j(0)$ and $\sigma^k x_j$ by $x_{j,k}=x_j(k)$, $k\geq 1$. 
\item Second evaluate $x_j(t)$ to numerical values by substituting the data.  
\end{itemize}
We will assume that we perform both steps in one: we will use $\sigma$ to denote $\sigma x_j(t)=x_j(t+1)$, $j=1,\ldots ,n$, to go from the value of the state variable at time $t$ to the value  at time $t+1$.

\subsubsection{Case of $n=1$} Let us assume we have only one state variable $x_1$ and $x_1(t+1)=\sigma x_1(t)$, so we consider
\begin{equation}
    F_1:=x_1(t+1) - p_1(x_1(t),\bm{\mu}),
\end{equation}
where $p_1$ is a polynomial in $\mathbb{C}[\bm{\mu}]$. 
We will now establish some notation. At $t=0$, we can write
\begin{equation}
    F_{1,0}=x_1(1)-\sum_{j=1}^s c_j(x_1(0)) M_j=x_{1,1}-\sum_{j=1}^s c_j(x_{1,0}) M_j
\end{equation}
where $M_j$ is a monomial in the variables $\bm{\mu}$,
with nonzero coefficients $c_{1,j}=c_{j}(x_1(0))\in \mathbb{C}$. 
\begin{definition}
The algebraic support of $F_{1,0}$ or $p_1$, as a polynomial in $\mathbb{C}[\bm{\mu}]$, is the following finite subset of $\mathbb{N}^L$: 
\[\supp  (F_{1,0})=\supp  (p_1)=\{\alpha\in \mathbb{N}^L\mid \bm{\mu}^{\alpha}\in \{M_1,\ldots ,M_s\}\},\]
where $\alpha=(\alpha_1,\ldots ,\alpha_L)$ and $\bm{\mu}^{\alpha}=\mu_1^{\alpha_1}\cdot\ldots \cdot \mu_L^{\alpha_L}$.
\end{definition}
Now apply   $k$ times $\sigma$ to $F_{1,0}$ to obtain $F_{1,k}:=\sigma^k F_{1,0}$. Let us consider the extended system 
\begin{equation}
     \sigma\Sigma'=\left\{F_{1,0},\ F_{1,1}:=\sigma F_{1,0},\ldots ,\ F_{1,L-1}=\sigma^{L-1} F_{1,0}\right\}.
\end{equation}

\begin{remark}\label{lemma:support}
Since  $\sigma c_j(x_{1,k})=c_j(x_{1,k+1})$, the extended system can be written as
\[
  \sigma\Sigma':%
  \begin{cases}
   F_{1,0} &= {x_{1,1}}- p_1(x_{1,0},\bm{\mu})=x_{1,1}-\sum_{j=1}^s c_j(x_{1,0}) M_j,\\
    &\vdots \\
    F_{1,L-1}  &= {x_{1,L}}-p_1(x_{1,L-1},\bm{\mu})=x_{1,L}-\sum_{j=1}^s c_j(x_{1,L-1}) M_j.
  \end{cases}
\]
Hence, $\sigma \Sigma'=\{F_{1,0},\ldots ,F_{1,L-1}\}$ is a system of $L$ polynomials in $\mathbb{C}[\bm{\mu}]$, recall $\bm{\mu} = (\mu_1,\ldots , \mu_L)$. Moreover, all  polynomials $F_{1,k}$ have the same support equal to $\supp  (F_{1,0})$.
\end{remark}

We define
\begin{align*}c_{1,j}&=c_{j}(x_{1,0})\\ c_{k+1,j}&=\sigma c_{j}(x_{1,k}), k\geq 1.
\end{align*}
Let  $C=(c_{i,j})$ denote the $L\times s$ coefficient matrix of the extended system $\sigma \Sigma'$ in the monomials  $\{M_1,\ldots ,M_s\}$. 
To study if the system 
\begin{equation}\label{eq:S_F}
    \mathcal{S}_{F_1}=\{F_{1,0}=0,\ldots ,F_{1,L-1}=0\}
\end{equation}
is consistent in $\bm{\mu}$, which means that it has a solution in $\mathbb{C}^L$, the coefficient matrix $C$ and $\supp (F_{1,0})$ have to be analyzed. In what follows, we illustrate
this analysis with examples. 

\begin{remark}
Complex solutions are usually not relevant to discrete-time models, so the complex solution set has to be analyzed afterwards to filter appropriate real solutions.
\end{remark}

\begin{example}\label{rem:5}
 The case $L=2$ can be reformulated as the intersection of two plane algebraic curves with the same support. For toy examples take:  
 \begin{enumerate}
     \item $F_{1,0}=x_{1,1}+c_{1,1}\mu_1^2+c_{1,2}\mu_2^2=0$ and  $\sigma F_{1,0}=x_{1,2}+c_{2,1}\mu_1^2+c_{2,2}\mu_2^2=0$. The system $\mathcal{S}_{F_1}$ has a finite number of solutions if  the coefficient matrix $C=(c_{i,j})$ is full rank. Additionally, the Jacobian matrix of $F_{1,0}, \sigma F_{1,0}$ w.r.t. $\mu_1,\mu_2$ is
     \[
     J = \begin{pmatrix}
     2c_{1,1}\mu_1 & 2c_{1,2}\mu_2\\
     2c_{2,1}\mu_1 & 2c_{2,2}\mu_2
     \end{pmatrix}.
     \]
     We have that $\det J = 4\mu_1\mu_2\cdot\det C$. Therefore, the consistency of $\mathcal{S}_{F_1}$ is equivalent to the non-degeneracy of $J$ in this case.

     \item $F_{1,0}=x_{1,0}+c_{1,1}\mu_1 \mu_2=0$ and  $\sigma F_{1,0}=x_{1,2}+c_{2,1}\mu_1 \mu_2=0$. The coefficient matrix is full rank, equal to $1$ and we have two parameters.  The system has either no solutions or infinitely many, a curve. Consider now the Jacobian matrix of $F_{1,0},\sigma F_{1,0}$:
     \[
     J = \begin{pmatrix}
     c_{1,1}\mu_2  & c_{1,1}\mu_1\\
     c_{2,1}\mu_2 & c_{2,1}\mu_1
     \end{pmatrix}.
     \]
     We then have $\det J = 0$, and so the inconsistency of $\mathcal{S}_{F_1}$ is equivalent to the degeneracy of $J$ for any coefficient matrix $C$ .
 \end{enumerate}
   \end{example}

\begin{example}[Ricker's Equation, \cite{Ricker}]\label{ex:Ricker}
Consider the equation, 
\begin{equation}
    N_{t+1}=N_t e^{r\left(1-\frac{N_t}{K}\right)}=N_t e^{\frac{r}{K}\left(K-N_t\right)}.
\end{equation}
with 
parameters $K$ and $rK:=\frac{r}{K}$. The output of Algorithm~\ref{alg:dynamictaylor} is the polynomial equation
\begin{equation}
    F:=N_{1}-N_0 p(N_0,K,rK), \mbox{ with } p=1+\sum_{j=0}^N \frac{rK^j(K-N_0)^j}{j!}.
\end{equation}
Let us study the solutions of the system $\mathcal{S}_F=\{F=0,\sigma F=0\}$, in the scalar parameters $K$ and $rK$. 
If $N_0>0$, $N_1>0$ and $N_0\neq N_1$, then the resultant $R$ of $F_1$ and $\sigma F_1$ with respect to $K$ is a nonzero polynomial in $\mathbb{C} [rK]$, since the leading coefficient is divisible by $N_0 N_1 (N_0 - N_1)$, which can be shows by a direct computation. In addition, if $N_0 N_2 - N_1^2\neq 0$ then $R$ has a nonzero solution. By the Extension Theorem then the system $\mathcal{S}_F$ is consistent for any degree $N$ of Taylor polynomial approximation.
A direct calculation also shows that
the Jacobian of $F_1$ and $\sigma F_1$ in the variables $K$ and $rK$ has leading coefficient $N_0 N_1 (N_0 - N_1)$ w.r.t. graded lex order. Therefore, under the same condition as above, $N_0 N_1 (N_0 - N_1)\ne 0$, the Jacobian is non-degenerate. 
\end{example}

So, in Example~\ref{ex:Ricker} (see also Example~\ref{rem:5}), we have the consistency of $\mathcal{S}_F$ and non-degeneracy of the Jacobian at the same time. We will use the latter as a heuristic test for consistency to speed up our algorithm in general. We will also prove, as in Example~\ref{ex:Ricker}, that for the series of examples we consider, independently of the degree of Taylor approximation, the non-degeneracy of the Jacobian  is a necessary condition for the consistency of the polynomial system.

\subsubsection{Case of $n > 1$, typical in applications: disjoint parameter sets.}\label{sec:disjoint} We will analyze first the scalar parameters $\bm{\mu}=(\mu_1,\ldots ,\mu_L)$ that appear in each polynomial equation of $\bm{F}=(F_1,\ldots , F_n)$. For this, we define the matrix $S(\Sigma)=(s_{i,j})$ as follows:
\[s_{i,j}:=\begin{cases}
    \alpha_{i,j} & \mbox{ if } \mu_j \mbox{ appears in } F_i,\\
    0 & \mbox{otherwise}.
\end{cases}\]
Let us assume that $n < L$ and study the shift of $\sigma\Sigma'$ to obtain a square system of $L$ equations in $L$ unknowns.
The number of shifts needed to make the system square is $L-n$. Let us denote the shifted system by $\sigma\Sigma'$.

We will consider next the  special case, occurring in biological models, in which the rows of $\sigma\Sigma'$ have disjoint sets of nonzero columns. This characterizes a system $\Sigma'$ in which each equation $F_i$ has an independent set $S_i$ of parameters.
Say $n=2$, $L=3$, $\bm{F}=(F_1,F_2)$ and
\[S(\Sigma)=\left(\begin{array}{ccc}
           \alpha_{1,1}  & \alpha_{1,2} & 0\\
           0  & 0 &\alpha_{2,3}
        \end{array}\right).\]
Then $F_1$ has parameter set $S_1=\{\mu_1,\mu_2\}$ and $F_2$ has parameter set $S_2=\{\mu_3\}$.
In this case, we define the shift by $\sigma\Sigma'=\{F_1,\sigma F_1, F_2 \}$.
The following statement follows immediately.

\begin{lemma}\label{lemma:disjoint}
    If the parameter sets $S_i$, $i=1,\ldots ,n$, are pairwise disjoint, then 
\begin{equation}
\sigma\Sigma'=\bigcup_{i=1}^n \{F_i,\sigma F_i,\ldots ,\sigma^{|S_i|-1} F_i\}
\end{equation}
is a set of $L$ polynomials equations in $L$ unknowns. Furthermore each subset $\{F_i,\sigma F_i,\ldots ,\sigma^{|S_i|-1} F_i\}$ has $|S_i|$ equations and $|S_i|$ unknowns.
\end{lemma}

To estimate the parameters $\bm{\mu}$, the system to be solved is
\[\mathcal{S}_{\Sigma'}=\bigcup_{i=1}^n \mathcal{S}_{F_i} \]
with $\mathcal{S}_{F_i}=\left\{F_i=0,\ \sigma F_i=0,\ldots ,\sigma^{|S_i|-1} F_i=0\right\}$. By Lemma \ref{lemma:disjoint}, the existence of solutions depends on the existence of solutions of each subsystem  $\mathcal{S}_{F_i}$.

\begin{example}[LPA Model]
We will now demonstrate our approach with the LPA model, which will bring a subtlety. Given the new model $B'$ 
 as the output of Algorithm~\ref{alg:dynamictaylor},  where  \[\exp(-c_{EL} L_t-c_{EA} A_t)\quad\text{and}\quad \exp(-c_{PA} A_t)\] have been approximated by
the polynomials
\begin{equation}
\hat{T}_{L_t}:=\sum_{k=0}^N \frac{1}{k!} (-(c_{EL}L_{t}+c_{EA}A_{t}) -\beta_t)^k\quad \text{and}\quad
\hat{T}_{A_{t}}:=\sum_{k=0}^N \frac{1}{k!}  (-c_{PA} A_{t}-\gamma_t)^k 
\end{equation}
respectively, we study now how to shift the equations in 
\begin{equation}
B_0:=\begin{cases}
F_{1,0}:=-L_{1} + b\cdot A_0\cdot \hat{T}_{L_0},\\
F_{2,0}:=-P_{1} + (1-\mu_L)\cdot L_0,\\
F_{3,0}:=-A_{1} + P_0\cdot \hat{T}_{A_0}+ (1-\mu_A)\cdot A_0,
\end{cases}
\end{equation}
to obtain a consistent square polynomial system.
Recall that the elements of $B_0$ are polynomials in $\mathbb{C}[a,b,d,c_{EL},c_{EA},c_{PA}]$.
For this, let us codify in a matrix $S(B_0)$ the indeterminates that appear in every equation. 
\[
\begin{array}{cc}
  &  \begin{array}{cccccc}
      b\quad & c_{EL}\quad & c_{EA}\quad & \mu_L\quad & c_{PA}\quad & \mu_A\quad
  \end{array} \\
  \begin{array}{c}
       F_{1,0}\\
       F_{2,0}\\
       F_{3,0}
  \end{array}   &  
\left( \begin{array}{cccccc}
    \alpha_{1,1} & \alpha_{1,2} & \alpha_{1,3} & 0 & 0 & 0 \\
      0 & 0 & 0 & \alpha_{2,4} & 0 & 0 \\
      0 & 0 & 0 & 0 & \alpha_{2,5} & \alpha_{2,6} 
\end{array}\right)
\end{array}
\]
Observe that $B_0$ is a system of $3$ equations in $6$ unknowns with the peculiarity that the sets of unknowns in each equation are disjoint. Observe the inconsistency that we arrive at, see the end of Section~\ref{ex:workedexample}, if we include equation $F_{2,1}$ because it contradicts  
$F_{2,0}$. So, we have to approach this differently. The shifted system
\[ \sigma B_0:=\{F_{1,0},F_{1,1}:=\sigma F_{1,0}, F_{1,2}:=\sigma^2 F_{1,0}, F_{2,0}, F_{3,0}, F_{3,1}:=\sigma F_{3,0}\} \]
has now $6$ equations in $6$ unknowns. The elements of $\sigma B_0$ are polynomials in $\mathbb{C}[b,\mu_L,\mu_A,c_{EL},c_{EA},c_{PA}]$ with the special property that a polynomial $F$ in $B_0$ and its shift $\sigma F$ have the same support. We will guarantee in Theorem \ref{theorem:jacobiansolvelpa} that the system of equations 
\begin{equation}
 \mathcal{S}_{LPA}=\{F_{1,0}=0,F_{1,1}=0, F_{1,2}=0, F_{2,0}=0, F_{3,0}=0, F_{3,1}=0\}   
\end{equation}
is consistent, which means that $\mathcal{S}_{LPA}$ has a common solution in $\mathbb{C}^6$. Complex solutions are not relevant to the LPA model, so the complex solution set has to be analyzed afterwards to filter appropriate real solutions. 

\begin{theorem}\label{theorem:jacobiansolvelpa}
Given system $B$ with nonzero initial conditions $L_0,P_0,A_0$, then the system $\mathcal{S}_{LPA}$, provided by the shifted set $\sigma B_0$,
is a consistent system of equations if and only if the following conditions hold:
\begin{enumerate}
    \item $F:=A_0 F_{3,1}- A_1 F_{3,0}$ is a non constant polynomial in $\mathbb{C}[c_{PA}]$.

    \item $G_1:=-L_1 Q_1+L_2 Q_0$ and  $G_2:=-L_1 Q_2+L_3 Q_0$, with $Q_t:=A_t \hat{T}_{L_t}$, are non-constant polynomials in $\mathbb{C}[c_{EL}, c_{EA}]$ defining curves with nonempty intersection.
\end{enumerate}
\end{theorem}
\begin{proof}
 Let $N$ be the degree of the polynomial approximations obtained in Algorithm~\ref{alg:dynamictaylor}. 
Since the sets of unknowns in $F_{1,0}$, $F_{2,0}$ and $F_{3,0}$ are disjoint, the shift subsets of each one of them give three subsystems with disjoint sets of unknowns. Proving that each one is consistent independently implies that the system defined by $\sigma B_0$ is consistent. 
More precisely:
\begin{enumerate}
    \item  $F_{2,0}=-P_{1} + b\cdot L_0$, under the assumption  $L_0\neq 0$ then $b$ is determined.

    \item $F_{3,0}$ and $F_{3,1}$ are  both polynomials with the same support
\begin{equation}
    F_{3,t}=-A_{t+1}+A_t (1-\mu_A) + \sum_{k=0}^N \frac{P_t}{k!}(-A_t c_{PA}-\gamma_t)^k,
    \quad t=0,1.
\end{equation}

$F=A_0 F_{3,1}- A_1 F_{3,0}$ is a polynomial in $\mathbb{C}[c_{PA}]$. 
Let $I=(F_{3,0}, F_{3,1})$ be the ideal generated by in $\mathbb{C} [\mu_A,c_{PA}]$.
If $F\in \mathbb{C}$  then $1\in I$, otherwise observe that $\{F_{3,0}, F\}$ is a Gr\"obner basis of $I$ w.r.t. the lexicographic monomial ordering with $\mu_A>c_{PA}$, since its leading monomials are pairwise relatively prime, namely $LM(F_{3,0})=\mu_A$ and $LM(F)=c_{PA}^J$, with $0\leq J\leq N$. Since $1$ does not belong to the ideal generated by $\mu_A$ and $c_{PA}^J$, we can guarantee that $1$ does not belong to the ideal $I$. 

    \item $F_{1,0}$, $F_{1,1}$  and $F_{1,2}$ are also polynomials with the same support
\[F_{1,t}=-L_{t+1}+ b \left(\sum_{k=0}^N \frac{A_t}{k!} (-L_t c_{EL}-A_t c_{EA}-\beta_t)^k\right)=-L_{t+1}+b Q_t,\quad t=0,1,2.\]
Let us define
\begin{equation}
    \begin{cases}
G_1:=Q_1 F_{1,0}- Q_0 F_{1,1}=-L_1 Q_1+L_2 Q_0,\\ 
G_2:=Q_2 F_{1,0}- Q_0 F_{1,2}=-L_1 Q_
2+L_3 Q_0. 
    \end{cases}
\end{equation}
Since $G_1, G_2\in \mathbb{C}[c_{EL}, c_{EA}]$, 
a common solution $(s_1,s_2)\in \mathbb{C}^2$ of $G_1(c_{EL},c_{EA})=0$ and $G_2(c_{EL},c_{EA})=0$ is an intersection point of the curves they define, in other words the algebraic variety $V(G_1,G_2)$ is not empty. 

If $G_1$ or $G_2$ belong to $\mathbb{C}$, or their resultant with respect to $c_{EL}$ (or $c_{EA}$) belongs to $\mathbb{C}$, then $1$ belongs to the ideal $J=(F_{1,0},F_{1,1},F_{1,2})$ of $\mathbb{C}[b, c_{EL}, c_{EA}]$. If $V(G_1,G_2)\subseteq V(Q_0)$ then $Q_0$ belongs to the ideal  $(G_1,G_2)$ of $\mathbb{C}[c_{EL}, c_{EA}]$, implying that \[-L_1=F_{1,0}-bQ_0\in J.\]
Otherwise, if $V(G_1,G_2)\not\subset V(Q_0)$, then $V(F_{1,0}, G_1, G_2)=V(F_{1,0}, F_{1,1}. F_{1,2})\ne \varnothing$.\qedhere 
\end{enumerate}
\end{proof}

\begin{remark}\label{rem:Jac}
The conditions stated in Theorem~\ref{theorem:jacobiansolvelpa} are equivalent to demanding the Jacobian matrix of the resulting polynomial system is symbolically of full rank. However, in this specific case, we can exhibit more details and present equations $F(c_{PA})=0$ and $\{G_1(c_{EL},c_{EA})=0$, $G_2(c_{EL},c_{EA})=0\}$ that would allow  us to compute complex solution sets for the parameters $c_{PA}$ and the pairs $(c_{EL},c_{EA})$, respectively. The discussion on how to filter real solutions continues in Algorithm~\ref{alg:squresystem}.
\end{remark}

\end{example}

\begin{example}[Discrete Lotka-Volterra models]
We now consider the discrete Lotka-Volterra model~\eqref{eq:DLVM}. We use Algorithm~\ref{alg:dynamictaylor} to obtain a polynomial system $\Sigma':\bm{F}=(F_1,\ldots ,F_n)$, where, for a fixed $i$,
\begin{equation}
    F_i:=x_i(t+1)-x_i(t) \sum_{k=0}^{N} \frac{p_i(t)^k}{k!}, \mbox{ where } p_i(t)=r_i-\sum_{j=1}^n a_{i,j} x_j(t).
\end{equation}
The parameter sets $S_i=\{r_i,a_{i,1},\ldots ,a_{i,n}\}$ are pairwise disjoint. To estimate the parameters in $S_i$ we solve the system 
\begin{equation}
\mathcal{S}_{F_i}=\{F_i=0,\sigma F_i=0,\ldots ,\sigma^n F_i=0\}.
\end{equation}
If we consider $x_j$ as variables, each equation is a dense polynomial in the variables $r_i$, $a_{i,j}$, $j=1,\ldots, n$, meaning that it contains all monomials up to order $N$. 
Thus studying the existence of solutions of the system $\mathcal{S}_{F_i}$
is 
a general problem of solving a multivariate polynomial system, most likely dense, where all the polynomials have the same degree $N$. 

The fact that the polynomials are generically dense allows us to give a result in terms of Macaulay resultants of multivariate polynomials.
Let us consider the resultant $R$ of
\begin{equation}
    \cF_i:=\{F_{i,0}:=F_i, F_{i,1}:=\sigma F_i,\ldots ,F_{i,n}:=\sigma^{n} F_i\}
\end{equation}
with respect to the variables $\cA_i:=\{a_{i,j}, j=1,\ldots , n\}$, defined as in \cite{DAndreaKrickSzanto}. In particular, it equals the Macaulay resultant $\res_{\cA_i}(\cF^h_i)$ of the homogenization $\cF^h_i:=\{F^h_{i,0}, F^h_{i,1},\ldots ,F^h_{i,n}\}$ of the polynomials in $\cF_i$ w.r.t. $\cA_i$, the classical projective resultant. It is a univariate polynomial $R(r_i)$ in $\mathbb{C} [r_i]$. Observe that $F_{i,j}=F^h_{i,j}(1,a_{i,1},\ldots ,a_{i,n})$, $j=0,\ldots ,n$ and 
set 
\begin{equation}
\overline{\cF}_i:=\{\overline{F}_{i,j}:=F^h_{i,j}(0,a_{i,1},\ldots ,a_{i,n}), j=0,1\ldots ,n-1\}.
\end{equation}

\begin{theorem}\label{thm:resultant}
    If $R$ is a non constant polynomial and $\res_{\cA_i}(\overline{\cF}_i)$ is non zero, then $\mathcal{S}_{F_i}$ has a solution in $\mathbb{C}^{n+1}$.
\end{theorem}
\begin{proof}
We apply formula (2) in \cite{DAndreaKrickSzanto} which reads 
\begin{equation}
    R(r_i)=\res_{\cA_i}(\cF^h_i)=\res_{\cA_i}(\overline{\cF}_i)^d \,\prod_{\xi\in V} F_{i,n}(\xi),
\end{equation}
where $d$ is the total degree of $F_{i,n}$ and 
\begin{equation}
    V=\{\xi\in \mathbb{C}^n \mid F_{i,1}(\xi)=0,\ldots , F_{i,n-1}(\xi)=0\}.
\end{equation}
\end{proof}

For instance, if $N=1$, we have linear systems
\begin{equation}
  \cF_i:%
  \begin{cases}
   F_{i,0} &= x_i(t+1)-x_i(t) (1+r_i-\sum_{j=1}^n a_{i,j} x_j(t)) ,\\
   F_{i,1} &= x_i(t+2)-x_i(t+1) (1 + r_i-\sum_{j=1}^n a_{i,j} x_j(t+1)),\\
   &\vdots \\
   F_{i,n} &= x_i(t+n+1)-x_i(t+n) (1 + r_i-\sum_{j=1}^n a_{i,j} x_j(t+n)).
  \end{cases}
\end{equation}
In this case, the resultant $R(r_i)$ is the determinant of the augmented matrix of $\cF_i$ in the variables $\cA_i$, a linear polynomial $a r_1+b$, where $a$ is the determinant of the coefficient matrix of $\cF_i$ in all the variables. 
If we denote $x_j(t+k)$ by $x_{j,k}$, then 
\begin{equation}
    \overline{F}_{i,k}=x_{i,k} \sum_{j=1}^n a_{i,j} x_{j,k}, k=0,\ldots ,n-1.
\end{equation}
we observe that $\res_{\cA_i}(\overline{\cF}_i)$ is the determinant of the coefficient matrix of $\overline{\cF}_i$. By Theorem \ref{thm:resultant}, we compute $r_i=s$ solving $R(r_i)=0$, then plug $r_i=s$ in $\overline{\cF}_i$ to compute the parameters $\cA_i$. 

If we leave $N$ arbitrary but fix $n=2$, we obtain
\begin{equation}
  \cF_i:%
  \begin{cases}
   F_{i,0} &= x_i(t+1)-x_i(t) \sum_{k=0}^{N} \frac{p_i(t)^k}{k!} ,\\
   F_{i,1} &= x_i(t+2)-x_i(t+1) \sum_{k=0}^{N} \frac{p_i(t+1)^k}{k!},\\
   F_{i,2} &= x_i(t+3)-x_i(t+2) \sum_{k=0}^{N} \frac{p_i(t+2)^k}{k!},
  \end{cases}
\end{equation}
where $p_i(t)=r_i-a_{i,1} x_1(t)-a_{i,2} x_2(t)$. Observe that 
\begin{equation}
    \overline{F}_{i,k}=-\frac{x_{i,k}}{N!} (-a_{i,1} x_{1,k}-a_{i,2} x_{2,k})^N,\quad k=0,1.
\end{equation}
Then $\res_{\cA_i}(\overline{\cF}_i)$ is the univariate resultant of $\overline{F}_{i,0}$ and $\overline{F}_{i,1}$ w.r.t. $a_{i,2}$ after replacing $a_{i,1}$ by $1$. More precisely,
\begin{equation}\label{eq:cond1}
    \res_{\cA_i}(\overline{\cF}_i)=\left(x_{i,0}x_{i,1}(x_{1,0}x_{2,1}-x_{1,1}x_{2,0})^N\right)^N.
\end{equation}
We do not give a general formula for $R(r_i)=\res_{\cA_i}(\cF^h_i)$ but we would like to emphasize that, for generic dense systems of polynomials, $R(r_i)$ gives effective conditions to guarantee that the system $\mathcal{S}_{F_i}$ is consistent.
For instance, for $n=2$ and $N=2$, we have 
\begin{equation}
    R(r_1)=\res_{\cA_1}(\cF^h_1)= \sum_{\ell=0}^8 \Gamma_{\ell} r_1^{\ell},
\end{equation}
where every $\Gamma_{\ell}$ is a nonzero polynomial in $\mathbb{Q}[x_{1,0}, x_{1,1}, x_{1,2},x_{1,3},x_{2,0}, x_{2,1}, x_{2,2}]$, considering $x_{i,j}$ as generic variables, and
\begin{equation}\label{eq:cond2}
    \Gamma_8= \frac{1}{4096} \left(x_{1,0} x_{1,1}x_{1,2} (x_{1,0}x_{2,1}-x_{1,0}x_{2,2}-x_{1,1}x_{2,0}+x_{1,1}x_{2,2} +x_{1,2}x_{2,0}-x_{1,2}x_{2,1})^2\right)^4.
\end{equation}
Thus, under conditions~\eqref{eq:cond1} and~\eqref{eq:cond2} not equal to zero,  we can guarantee by Theorem \ref{thm:resultant} that the system $\mathcal{S}_{F_i}$ is consistent. Furthermore, 
\begin{equation}\label{eq:cond3}
     x_{1,0} x_{1,1}x_{1,2} (x_{1,0}x_{2,1}-x_{1,0}x_{2,2}-x_{1,1}x_{2,0}+x_{1,1}x_{2,2} +x_{1,2}x_{2,0}-x_{1,2}x_{2,1})
\end{equation}
is the independent term of the Jacobian of $\cF_i$ with respect to $r_i, a_{1,1}, a_{1,2}$ is non-zero. Thus having a non-zero Jacobian is again a necessary condition for the consistency of the system. 
These computations were performed with {\sc Maple} and are available   at \url{https://github.com/soniarueda/discrete-time-models-exp}. 
\end{example}

\subsubsection{Case of $n > 1$, general.}

If the hypothesis of Lemma \ref{lemma:disjoint} does not hold, the parameter sets are not disjoint, then we will duplicate the common parameters to force disjoint parameter sets.
Given the system $\Sigma'$ in \eqref{eq:Sigmap} we consider a new system 
\begin{equation}
  \Sigma^e:  \bm{E} = {\bm{x}_{t+1}}- \bm{q}(\bm{x}_t,\bm{\mu}^e)
\end{equation}
with $\bm{\mu}^e=(\mu_1^e,\ldots ,\mu_L^e)$, being $\mu_j^e=\mu_{j,1},\ldots ,\mu_{j,\ell_i}$, where $\ell_j$ is the number of equations from $\Sigma'$ where $\mu_j$ appears. 
The polynomials 
\[q_i=p_i(\bm{x}_t, \mu_{1,i},\ldots ,\mu_{L,i}), \mbox{ with } \mu_{j,i}=\begin{cases}
    \mu_j & \mbox{ if } \ell_j=1\\
    \mbox{ a new variable } & \mbox{ if } \ell_j>1
\end{cases}.\]
For instance if $n=2$, $L=3$ and
        \[S(\Sigma)=\left(\begin{array}{ccc}
           \alpha_{1,1}  & \alpha_{1,2} & \alpha_{1,3}\\
           \alpha_{2,1}  & 0 &0
        \end{array}\right)\]
then $\mu_1$ is a common parameter in $F_1$ and $F_2$, then $\bm{\mu}^e=(\mu_{1,1},\mu_{1,2},\mu_2 ,\mu_3)$ and $\Sigma^e$ is now a system of $n=2$ equations in $4$ parameters. 

In general, new system $\Sigma^e$ has $n$ polynomial equations $\bm{E}=(E_1,\ldots ,E_n)$ in $L^e=\ell_1+\cdots +\ell_L$ parameters $\bm{\mu}^e$. The new system $\Sigma^e$ satisfies now the hypothesis of Lemma \ref{lemma:disjoint}, and to estimate the parameters $\bm{\mu^e}$ we solve the system $\mathcal{S}_{\Sigma^e}$.

\begin{example}[Discrete Lotka-Volterra models]

  \begin{equation}
  \Sigma:%
  \begin{cases}
   x_{t+1} &=x_t  e^{r+a_{1,1}x_t-a_{1,2} y_t},\\
   y_{t+1} &=y_t e^{r+a_{2,1}x_t-a_{2,2} y_t}.
  \end{cases}%
\end{equation}
These two equations have the parameter $r$ in common. We use Algorithm~\ref{alg:dynamictaylor} to obtain a polynomial system $\Sigma'$. We consider the extended system $\Sigma^e$ which has now  distinct  parameters $r_1$ and $r_2$ and satiefies the hypotesis of Lemma   \ref{lemma:disjoint}
\begin{equation}
  \Sigma^e:%
  \begin{cases}
   E_1 &:= x_{t+1}-x_t p_1(x_t,y_t,r_1,a_{1,1},a_{1,2}) ,\\
   E_2 &:= y_{t+1}-y_t p_2(x_t,y_t,r_2,a_{2,1},a_{2,2}).
  \end{cases}%
\end{equation}
The shift is then a system
\begin{equation}
   \sigma\Sigma^e=\{E_1,\sigma E_1,\sigma^2 E_1, E_2,\sigma E_2,\sigma^2 E_2\} 
\end{equation}
of $6$ equations in $6$ unknowns. Since the parameters sets in $E_1$ and $E_2$ are disjoint, we solve $\mathcal{S}_{E_1}$ and $\mathcal{S}_{E_2}$ independently.

\end{example}

\begin{algorithm}[H]
\caption{Square system method}
\label{alg:squresystem}
    \begin{algorithmic}[1]
        \State {\bf Input}:\begin{itemize}
            \item Algebraic system of difference equations named $\Sigma'$ with some of the coefficients given as intervals
            \item Time-measured data allowing shift of the system.
            \item For $\bar{\mu}=\mu_1,\ldots,\mu_n$ the finite set of parameters, the data $R_{\mu_i}$ of permissible intervals for each parameter value.
        \end{itemize}
        \State {\bf Output}: Parameter values of $\Sigma'$.
    \Procedure{Detect solvability}{} 
        \State Define $\Sigma_\mathrm{shift}$ as an indefinite time shift of $\Sigma'$.
        \State Redefine $\Sigma':=[]$, an `empty system' of no equations.
        \State Denote $e_1,\ldots$ the equations of $\Sigma_\mathrm{shift}$.
        \State Define $J=\left[\dfrac{\partial e_i}{\partial\mu_j}\right]$ where $e_i\in \Sigma',i=1,\ldots,r$, or 0 if $\Sigma'$ is empty.
        \State $i:=1$
        \For{$\rank(J):=s<n$ i.e. is not of full rank, }
            \Procedure{Move equation}{}
            \State \If {data are available, pick new $e_i$}  
            \State $\Sigma':=\Sigma'\cup\{e_i\}$
            \State $\Sigma_\mathrm{shift}:=\Sigma_\mathrm{shift}\backslash\{e_i\}$
            \Else{ return failure}\EndIf
            \State Compute $\rank(J)$ (note $\Sigma'$ has updated)
            \If{$\rank(J)<s+1$} $\Sigma':=\Sigma'\backslash\{e_i\}$
            \EndIf
            \State $i=i+1$
            \If{$\rank(J)<n$} repeat procedure Move Equation.
            \EndIf
            \EndProcedure
        \EndFor
    \Procedure{Blackbox solver and filter}{} Note Detect Solvability runs (as a \phantom{---------}heuristic) until the Jacobian matrix $J$  has full rank and outputs a polynomial system \phantom{----------}$\Sigma'$  that has  solutions.  Run any polynomial system solver that accepts coefficients as \phantom{---------} intervals (see Remark~\ref{rem:4}).
        \State Filter solutions by intersecting solution set with $R_{\mu_i}$.
        \EndProcedure
        \EndProcedure
    \end{algorithmic}
\end{algorithm}

\begin{remark}
The procedure `Move Equation' extracts equations from the shift and adds them to $\Sigma'$, without replacement. It checks if this increases the rank of the Jacobian. If it does not, it deletes the equation from $\Sigma'$ as well, removing it entirely from consideration. The process repeats until the Jacobian reaches full rank in the number of parameters $n$.
\end{remark}

\begin{remark}\label{rem:11}
Our algorithm uses the full-rank Jacobian condition as a heuristic to detect solvability. It must be noted that full rank Jacobian is not a sufficient condition to have solutions in the general setting. For instance the system $\{xy=1,x=0\}$ has no solutions but has symbolically full-rank Jacobian
\[
\begin{pmatrix}
y & x\\
1 & 0
\end{pmatrix}
.\] However, in Remark~\ref{rem:Jac}, we see that, for instance, for the LPA system, the corresponding polynomial system does have solutions if and only if the Jacobian of is of full rank. It would be interesting to expand classes of difference models for which such a criterion still holds, probably extending the proof of Theorem~\ref{theorem:jacobiansolvelpa}.
\end{remark}

\subsection{Continued example: getting consistency and filtering multiple solutions}\label{ex:continuedexample}

We continue from our earlier example, Section~\ref{ex:workedexample}. Removing the uninteresting first few equations, we have:
\begin{equation}\label{eq:examplesystem:rem}
    B':\begin{cases}
       L_1=33=214b\left(0.18 + 0.04(-107c_{EL}-214c_{EA})\right. \\ \qquad\qquad\left.+ 0.02(-107c_{EL}-214c_{EA}+ 3.17)^2\right)\\
        P_1=86=107(1-\mu_{L})\\
        A_1=239=73\left(0.69 + 0.32(-214c_{PA})\right. \\ \qquad\qquad\left.+ 0.16(-214c_{PA} + 1.13)^2\right)+214\cdot(1-\mu_A)
        \\
        L_2=67 = 240b(0.1 - 0.67c_{EL} - 4.89c_{EA}+ 0.01(-33c_{EL} - 240c_{EA} + 3.89)^2)\\
        P_2=27= 33(1-\mu_{L})\\
        A_2=267 = 55.29 - 5884.55c_{PA}  + 12.26(-240c_{PA} + 1.25)^2+ 240(1-\mu_A)
    \end{cases}
\end{equation}
This forms 6 equations in 6 unknowns. Incidentally, it is immediately obvious that the 2nd and 5th equations force the system to be inconsistent. A system will rarely have such an obvious inconsistency however, and we demonstrate that the method of the Jacobian yields the same answer. By computing the Jacobian and converting it to Reduced Row-Echelon Form (RREF) we find:
\begin{equation*}
    J\overset{RREF}{=}\begin{bmatrix}
        1&0&0&0&0&0\\
        0&1&0&0&*&0\\
        0&0&1&0&0&0\\
        0&0&0&1&*&0\\
        0&0&0&0&0&1\\
        0&0&0&0&0&0
    \end{bmatrix}
\end{equation*}
where the $*$ are (non-equal) large, nonzero algebraic expressions in the parameters. As our interest is in the rank, we omit the expressions for readability. It is clear that the Jacobian has rank 5 and so an equation must be replaced. Note the Jacobian is written in an ordered basis of the parameters, specifically $\mu_L,b,\mu_A,c_{EL},c_{EA},c_{PA}$.

\begin{remark} In general, the approach via computing the Jacobian is a heuristic - see Remark~\ref{rem:11}. So, a reliable way to test the consistency is still to try to solve the polynomial system. The rank computation here was done via RREF for illustration purposes only. In practice, for larger problems, it would be more reasonable to do a numerical rank computation by random substitution into the symbolic variables to get a lower bound for the rank.
\end{remark}

In practice, the choice of which equation to replace can be made by determining which row (equivalently, equation) is a linear combination of the others. 
Here, before row reducing, one finds row 2 is $33/107$ times row 5, so either may be deleted. We choose to delete the 5th equation. We will need to shift the system using time-measured data. Suppose this is available (i.e., the scientist has continued the experiment) and is recorded in Table~\ref{tab:continuedexampledata}, with Algorithm~\ref{alg:expansionrange} using Table~\ref{tab:acceptablerangesexample} and $t=1,2$ data as in Section~\ref{ex:workedexample}.
\begin{table}
    \centering
    \begin{tabular}{|c|c|c|c|}\hline
         $t$& $L_t$ &$P_t$&$A_t$\\\hline
         3& 36 & 54&273\\\hline\hline
         $t$&$R_{L_t}$&-&$R_{A_t}$\\\hline
         3&$[-4.16,-3.02]$&-&$[-1.69,-0.87]$\\\hline
         &$\alpha=-3.59$&-&$\beta=-1.28$\\\hline
    \end{tabular}
\caption{
  Shift data and results from Algorithm~\ref{alg:expansionrange}}
\label{tab:continuedexampledata}
\end{table}
Applying Algorithm~\ref{alg:dynamictaylor} up to degree 2 as before yields the new equations:
\begin{equation*}
    \begin{cases}
        L_3=36 = 267b(0.13 - 1.85c_{EL} - 7.38c_{EA} + 0.014(-67c_{EL} - 267c_{EA} + 3.59)^2)\\
        P_3=54 = 67(1-\mu_L)\\
        A_3=273 = 17.09 - 1998.17c_{PA} + 3.74(-267c_{PA} + 1.28)^2 + 267(1-\mu_A) 
    \end{cases}
\end{equation*}
Next, we decide which of these new equations to replace the former equation 5. In practice, we implement by selecting `down the line' with the first available equation, checking  for consistency, and proceeding until solutions are found. Of course, there are smarter options here, one notes for instance $21=26(1-\mu_L)$ is a bad choice as this creates 
 an inconsistent system; but in highly complicated systems, we cannot expect such obviousness.

If we replace the equation from $P_2$ with the computed $L_3$, we turn to a solver to find:

\begin{equation}\label{eq:examplesystem:ext}
    B':\begin{cases}
       L_1=33=214b\left(0.18 + 0.04(-107c_{EL}-214c_{EA})\right. \\ \qquad\qquad\left.+ 0.02(-107c_{EL}-214c_{EA}+ 3.17)^2\right)
    \\P_1=85=107(1-\mu_{L})\\
        A_1=239=73\left(0.69 + 0.32(-214c_{PA})\right. \\ \qquad\qquad\left.+ 0.16(-214c_{PA} + 1.13)^2\right)+214\cdot(1-\mu_A)
        \\
        L_2=67 = 240b(0.1 - 0.67c_{EL} - 4.89c_{EA} + 0.01(-33c_{EL} - 240c_{EA} + 3.89)^2)\\
         L_3=36 = 267b(0.13 - 1.85c_{EL} - 7.38c_{EA} + 0.01(-67c_{EL} - 267c_{EA} + 3.59)^2)\\
        A_2=267 = 55.29 - 5884.55c_{PA} + 12.26(-240c_{PA} + 1.25)^2 + 240(1-\mu_A)
    \end{cases}
\end{equation}
Using {\sc Maple}, we received the multiple solutions:
\begin{gather*}
        \mu_L = 0.196, b = 6.823, \mu_A = 0.002, c_{EL} = 0.014, c_{EA} = 0.011, c_{PA} = 0.005\\ \mu_L = 0.196, b = 0.048, \mu_A = 0.002, c_{EL} = -0.434, c_{EA} = 0.179, c_{PA} = 0.005\\ \mu_L = 0.196, b = 10.225+ 1.969i, \mu_A = 0.002, c_{EL} = 0.011 + 0.005i, c_{EA} = 0.014 + 0.0001i, c_{PA} = 0.005\\ \mu_L = 0.196, b = 10.225 - 1.969i, \mu_A = 0.002, c_{EL} = 0.011 - 0.005i, c_{EA} = 0.014 - 0.0001i, c_{PA} = 0.005\\ \mu_L = 0.196, b = 6.823, \mu_A = 0.047, c_{EL} = 0.014, c_{EA} = 0.011, c_{PA} = 0.007\\ \mu_L = 0.197, b = 0.048, \mu_A = 0.047, c_{EL} = -0.434, c_{EA} = 0.179, c_{PA} = 0.007\\ \mu_L = 0.196, b = 10.225 + 1.969i, \mu_A = 0.047, c_{EL} = 0.011 + 0.005i, c_{EA} = 0.0136 + 0.0001i, c_{PA} = 0.007\\ \mu_L = 0.196, b = 10.225 - 1.969i, \mu_A = 0.047, c_{EL} = 0.011 - 0.005i, c_{EA} = 0.0136 - 0.0001i, c_{PA} = 0.007
\end{gather*}
To discriminate which answer to use, we intersect with the acceptable intervals for the parameters $c_{EL},c_{EA},c_{PA}$ given in Section~\ref{ex:workedexample}. (Note then, as mentioned, complex and negative real solutions never matter). Here this returns the correct answer as the first provided solution. Hence our algorithm estimates:
\[  \mu_L = 0.196,\ \ b = 6.823,\ \ \mu_A = 0.002,\ \ c_{EL} = 0.014,\ \ c_{EA} = 0.011,\ \ c_{PA} = 0.005.
\]
This experiment was run as a simulation so we know the actual parameters were:
\[
                                \mu_L= 0.206,\ \
                         b= 6.598,\ \
                      \mu_A= 0.007629,\ \
                     c_{EL}= 0.01209,\ \
                     c_{EA}= 0.01155,\ \
                     c_{PA}= 0.0047.
\]
In Section~\ref{sect:experiment} we run this experiment repeatedly to measure statistical data on Algorithm~\ref{alg:squresystem}'s success. Full details are provided there, but to summarize, it works well.

\section{Performance analysis}\label{sect:experiment}

In this section, we demonstrate the quality of our method  is demonstrated in two experiments. The first is  by applying it to the LPA model. The code is available at \url{https://github.com/jedforrest/larva-pupa-adult/}. 
The results of this experiment are in Tables~\ref{tab:numericalresults1} and~\ref{tab:numericalresults2}.  The second is by applying to the competition model as in~\eqref{eq:DLVM}, with code available at \url{https://github.com/alexeyovchinnikov/discrete-time-models-exp/}. 

\subsection{Experiment setup}
To test the algorithm, we begin by choosing explicit baseline values for each of the 6 parameters $\{b,c_{EL},c_{EA},c_{PA},\mu_L,\mu_A\}$. For each run of the experiment, a set of `ground truth' parameters $\mathbf{\hat{x}}$ are randomly sampled around the baseline values. In particular, starting from the experimentally determined values given in \cite{CDCD} we generated small intervals $I_1, \ldots, I_6$ centered around each of the values in \cite{CDCD}, then uniformly sampled parameters from the intervals.\footnote{Use of \cite{CDCD} is for scientific relevance to LPA, however this knowledge of existing estimates is not necessary when choosing a baseline.} These sampled parameters are used to generate $N$ sets of synthetic time measured data for the larva $L_t$, pupa $P_t$, and adult $A_t$ populations over 3 shifts. Synthetic data is generated using an exact $\exp$ function.

For a selected degree $N$, our systems of polynomials are generated with Taylor polynomial degree $N$ according to the earlier defined Algorithm~\ref{alg:dynamictaylor}. These equations take the synthetic data and solve for the parameters using HomotopyContinuation.jl~\cite{HomotopyContinuation.jl}. The shifts for each variable are handled separately so that we have 3 sets of polynomials (one for each of $L,P,A$). If there are more equations than parameters to fit, a subset of equations is chosen equal to the number of parameters as in Algorithm~\ref{alg:squresystem}.

The solver finds real solutions for all parameters. If the solver returns multiple real solutions, then they are filtered using the defined intervals $I$ so that our solutions exist in the correct domain. If no real solution is found for a particular parameter within the interval, the algorithm returns $0$ for that parameter (indicating an unsuccessful search).

We used the following experimental settings:
\begin{itemize}
    \item Interval ranges: $I\in\{\pm5\%,\pm10\%,\pm20\%,\pm25\%,\pm50\%\}$
    \item Taylor degrees: $N\in\{1, 2, \ldots, 10\}$
    \item Number of data sets generated: $D=100$
\end{itemize}
In total there were $|I|\cdot|N|\cdot|D|=5000$ simulations.

The mean and median of relative errors $\delta$ were calculated across all $D$ simulated data sets, grouped by interval range $I$ and Taylor degree $N$. The relative error for each parameter $j$ is defined as
\begin{equation*}
    \delta_j = \left|\frac{x_j-\hat{x_j}}{\hat{x_j}}\right|
\end{equation*}
where $x_j$ is the predicted parameter value for parameter $j$ and $\hat{x_j}$ is the actual (true) parameter value.

\subsection{Presentation of the performance}

\subsubsection{Applied to LPA}
In Tables~\ref{tab:numericalresults1} and~\ref{tab:numericalresults2}, each labelled column presents the error in scientific notation rounded to three significant digits. Adjacent to the right, the same values are given as a percentage rounded to the nearest decimal percent to emphasize the general success of the method.

This  captures the general trend that increasing $N$ increases accuracy, and that while wider intervals generally lower the accuracy, this can be combated by further increasing $N$. It is not computationally prohibitive to achieve accuracy within 5 decimal places in all parameters. Note that because of the simplicity of $P_{t+1}=(1-\mu_L)L_{t}$ (a linear polynomial with no $\exp$), the parameter $\mu_L$ is immediately recovered.

The algorithm is progressively better with increasing degree of Taylor polynomial $n$, so much so that in the worst scenarios investigated, $n\ge4$ is sufficient to significantly minimize error. The method generally gets worse with longer intervals but this is easily overcome by increasing the degree of the Taylor polynomial.

We conclude that the method of replacing $\exp$ in the LPA model with algebraic equations to estimate parameters is highly successful. We further suggest that this method will continue to be highly successful in replacing $\exp$ in more general difference equation settings.

\newpage
\begin{table}[!ht]
    \centering
    \tiny
    \begin{tabular}{|l|l|l|l|l|l|l|l|l|l|l|l|l|l|}
    \hline
        $N$ & Interval & Mean $b$ &  & Med $b$ &  & Mean $c_{EA}$ &  & Med $c_{EA}$ &  & Mean $c_{EL}$ & & Med $c_{EL}$ &  \\ \hline
        \hline
        1 & $\pm5\%$ & 7.16E-02 & 7.2\% & 1.51E-03 & 0.2\% & 7.15E-02 & 7.1\% & 1.39E-03 & 0.1\% & 7.11E-02 & 7.1\% & 9.71E-04 & 0.1\% \\
        2 & $\pm5\%$ & 9.16E-05 & {\bf 0.0\%} & 4.59E-05 & {\bf 0.0\%} & 8.26E-05 & {\bf 0.0\%} & 4.75E-05 & {\bf 0.0\%} & 5.83E-05 & {\bf 0.0\%} & 2.93E-05 & {\bf 0.0\%} \\ 
        3 & $\pm5\%$ & 3.67E-06 & {\bf 0.0\%} & 1.36E-06 & {\bf 0.0\%} & 3.10E-06 & {\bf 0.0\%} & 1.17E-06 & {\bf 0.0\%} & 1.87E-06 & {\bf 0.0\%} & 5.79E-07 & {\bf 0.0\%} \\ 
        4 & $\pm5\%$ & 1.19E-07 & {\bf 0.0\%} & 2.31E-08 & {\bf 0.0\%} & 1.00E-07 & {\bf 0.0\%} & 2.09E-08 & {\bf 0.0\%} & 5.16E-08 & {\bf 0.0\%} & 9.96E-09 & {\bf 0.0\%} \\ 
        5 & $\pm5\%$ & 3.38E-09 & {\bf 0.0\%} & 3.76E-10 & {\bf 0.0\%} & 2.86E-09 & {\bf 0.0\%} & 3.31E-10 & {\bf 0.0\%} & 1.23E-09 & {\bf 0.0\%} & 1.30E-10 & {\bf 0.0\%} \\ 
        6 & $\pm5\%$ & 8.62E-11 & {\bf 0.0\%} & 5.56E-12 & {\bf 0.0\%} & 7.33E-11 & {\bf 0.0\%} & 4.36E-12 & {\bf 0.0\%} & 2.64E-11 & {\bf 0.0\%} & 1.71E-12 & {\bf 0.0\%} \\ 
        7 & $\pm5\%$ & 2.48E-12 & {\bf 0.0\%} & 8.59E-13 & {\bf 0.0\%} & 2.13E-12 & {\bf 0.0\%} & 7.74E-13 & {\bf 0.0\%} & 9.55E-13 & {\bf 0.0\%} & 4.96E-13 & {\bf 0.0\%} \\ 
        8 & $\pm5\%$ & 8.66E-13 & {\bf 0.0\%} & 7.33E-13 & {\bf 0.0\%} & 7.55E-13 & {\bf 0.0\%} & 6.57E-13 & {\bf 0.0\%} & 4.07E-13 & {\bf 0.0\%} & 4.04E-13 & {\bf 0.0\%} \\
        9 & $\pm5\%$ & 9.32E-13 & {\bf 0.0\%} & 7.75E-13 & {\bf 0.0\%} & 7.87E-13 & {\bf 0.0\%} & 6.35E-13 & {\bf 0.0\%} & 1.06E-13 & {\bf 0.0\%} & 1.07E-13 & {\bf 0.0\%} \\ 
        10 & $\pm5\%$ & 1.18E-12 & {\bf 0.0\%} & 1.03E-12 & {\bf 0.0\%} & 1.00E-12 & {\bf 0.0\%} & 8.61E-13 & {\bf 0.0\%} & 3.16E-13 & {\bf 0.0\%} & 3.16E-13 & {\bf 0.0\%} \\ \hline
        1 & $\pm10\%$ & 1.06E-01 & 10.6\% & 6.35E-03 & 0.6\% & 1.06E-01 & 10.6\% & 5.70E-03 & 0.6\% & 1.04E-01 & 10.4\% & 4.23E-03 & 0.4\% \\
        2 & $\pm10\%$ & 5.08E-02 & 5.1\% & 4.95E-04 & {\bf 0.0\%} & 5.07E-02 & 5.1\% & 4.55E-04 & {\bf 0.0\%} & 5.05E-02 & 5.0\% & 2.84E-04 & {\bf 0.0\%} \\ 
        3 & $\pm10\%$ & 8.09E-05 & {\bf 0.0\%} & 2.53E-05 & {\bf 0.0\%} & 6.67E-05 & {\bf 0.0\%} & 2.46E-05 & {\bf 0.0\%} & 3.97E-05 & {\bf 0.0\%} & 7.69E-06 & {\bf 0.0\%} \\ 
        4 & $\pm10\%$ & 5.84E-06 & {\bf 0.0\%} & 1.15E-06 & {\bf 0.0\%} & 4.77E-06 & {\bf 0.0\%} & 1.04E-06 & {\bf 0.0\%} & 2.42E-06 & {\bf 0.0\%} & 2.75E-07 & {\bf 0.0\%} \\ 
        5 & $\pm10\%$ & 3.63E-07 & {\bf 0.0\%} & 4.14E-08 & {\bf 0.0\%} & 2.95E-07 & {\bf 0.0\%} & 3.50E-08 & {\bf 0.0\%} & 1.26E-07 & {\bf 0.0\%} & 7.63E-09 & {\bf 0.0\%} \\
        6 & $\pm10\%$ & 2.01E-08 & {\bf 0.0\%} & 1.16E-09 & {\bf 0.0\%} & 1.63E-08 & {\bf 0.0\%} & 1.10E-09 & {\bf 0.0\%} & 5.88E-09 & {\bf 0.0\%} & 1.93E-10 & {\bf 0.0\%} \\ 
        7 & $\pm10\%$ & 1.00E-09 & {\bf 0.0\%} & 3.20E-11 & {\bf 0.0\%} & 8.13E-10 & {\bf 0.0\%} & 3.07E-11 & {\bf 0.0\%} & 2.47E-10 & {\bf 0.0\%} & 4.96E-12 & {\bf 0.0\%} \\ 
        8 & $\pm10\%$ & 4.62E-11 & {\bf 0.0\%} & 1.66E-12 & {\bf 0.0\%} & 3.73E-11 & {\bf 0.0\%} & 1.40E-12 & {\bf 0.0\%} & 9.75E-12 & {\bf 0.0\%} & 4.76E-13 & {\bf 0.0\%} \\ 
        9 & $\pm10\%$ & 2.67E-12 & {\bf 0.0\%} & 1.00E-12 & {\bf 0.0\%} & 2.18E-12 & {\bf 0.0\%} & 8.30E-13 & {\bf 0.0\%} & 3.72E-13 & {\bf 0.0\%} & 1.13E-13 & {\bf 0.0\%} \\ 
        10 & $\pm10\%$ & 1.58E-12 & {\bf 0.0\%} & 1.09E-12 & {\bf 0.0\%} & 1.35E-12 & {\bf 0.0\%} & 9.27E-13 & {\bf 0.0\%} & 3.24E-13 & {\bf 0.0\%} & 3.22E-13 & {\bf 0.0\%} \\ \hline
        1 & $\pm20\%$ & 2.67E-01 & 26.7\% & 2.60E-02 & 2.6\% & 2.66E-01 & 26.6\% & 2.28E-02 & 2.3\% & 2.61E-01 & 26.1\% & 1.75E-02 & 1.7\% \\ 
        2 & $\pm20\%$ & 2.35E-01 & 23.5\% & 5.48E-03 & 0.5\% & 2.34E-01 & 23.4\% & 4.88E-03 & 0.5\% & 2.32E-01 & 23.2\% & 2.47E-03 & 0.2\% \\ 
        3 & $\pm20\%$ & 3.14E-02 & 3.1\% & 4.30E-04 & {\bf 0.0\%} & 3.12E-02 & 3.1\% & 3.68E-04 & {\bf 0.0\%} & 3.06E-02 & 3.1\% & 1.20E-04 & {\bf 0.0\%} \\ 
        4 & $\pm20\%$ & 2.42E-04 & {\bf 0.0\%} & 3.90E-05 & {\bf 0.0\%} & 1.99E-04 & {\bf 0.0\%} & 3.18E-05 & {\bf 0.0\%} & 7.49E-05 & {\bf 0.0\%} & 7.86E-06 & {\bf 0.0\%} \\ 
        5 & $\pm20\%$ & 2.83E-05 & {\bf 0.0\%} & 3.06E-06 & {\bf 0.0\%} & 2.32E-05 & {\bf 0.0\%} & 2.85E-06 & {\bf 0.0\%} & 7.70E-06 & {\bf 0.0\%} & 4.75E-07 & {\bf 0.0\%} \\ 
        6 & $\pm20\%$ & 3.06E-06 & {\bf 0.0\%} & 2.43E-07 & {\bf 0.0\%} & 2.49E-06 & {\bf 0.0\%} & 2.20E-07 & {\bf 0.0\%} & 7.31E-07 & {\bf 0.0\%} & 2.34E-08 & {\bf 0.0\%} \\ 
        7 & $\pm20\%$ & 2.95E-07 & {\bf 0.0\%} & 1.31E-08 & {\bf 0.0\%} & 2.38E-07 & {\bf 0.0\%} & 1.23E-08 & {\bf 0.0\%} & 6.23E-08 & {\bf 0.0\%} & 1.02E-09 & {\bf 0.0\%} \\
        8 & $\pm20\%$ & 2.58E-08 & {\bf 0.0\%} & 6.75E-10 & {\bf 0.0\%} & 2.07E-08 & {\bf 0.0\%} & 5.45E-10 & {\bf 0.0\%} & 4.85E-09 & {\bf 0.0\%} & 4.89E-11 & {\bf 0.0\%} \\
        9 & $\pm20\%$ & 2.07E-09 & {\bf 0.0\%} & 2.72E-11 & {\bf 0.0\%} & 1.65E-09 & {\bf 0.0\%} & 2.20E-11 & {\bf 0.0\%} & 3.48E-10 & {\bf 0.0\%} & 1.70E-12 & {\bf 0.0\%} \\ 
        10 & $\pm20\%$ & 1.54E-10 & {\bf 0.0\%} & 3.00E-12 & {\bf 0.0\%} & 1.22E-10 & {\bf 0.0\%} & 2.50E-12 & {\bf 0.0\%} & 2.31E-11 & {\bf 0.0\%} & 3.96E-13 & {\bf 0.0\%} \\ \hline
        1 & $\pm25\%$ & 3.39E-01 & 33.9\% & 3.99E-02 & 4.0\% & 3.37E-01 & 33.7\% & 3.71E-02 & 3.7\% & 3.34E-01 & 33.4\% & 3.11E-02 & 3.1\% \\
        2 & $\pm25\%$ & 2.35E-01 & 23.5\% & 6.20E-03 & 0.6\% & 2.35E-01 & 23.5\% & 5.84E-03 & 0.6\% & 2.34E-01 & 23.4\% & 4.09E-03 & 0.4\% \\ 
        3 & $\pm25\%$ & 1.25E-02 & 1.2\% & 6.99E-04 & 0.1\% & 1.21E-02 & 1.2\% & 6.28E-04 & 0.1\% & 1.13E-02 & 1.1\% & 2.69E-04 & {\bf 0.0\%} \\ 
        4 & $\pm25\%$ & 2.05E-02 & 2.0\% & 6.54E-05 & {\bf 0.0\%} & 2.04E-02 & 2.0\% & 6.64E-05 & {\bf 0.0\%} & 2.02E-02 & 2.0\% & 2.06E-05 & {\bf 0.0\%} \\
        5 & $\pm25\%$ & 6.99E-05 & {\bf 0.0\%} & 4.89E-06 & {\bf 0.0\%} & 5.75E-05 & {\bf 0.0\%} & 5.50E-06 & {\bf 0.0\%} & 2.86E-05 & {\bf 0.0\%} & 1.24E-06 & {\bf 0.0\%} \\ 
        6 & $\pm25\%$ & 9.00E-06 & {\bf 0.0\%} & 4.32E-07 & {\bf 0.0\%} & 7.56E-06 & {\bf 0.0\%} & 3.94E-07 & {\bf 0.0\%} & 3.39E-06 & {\bf 0.0\%} & 6.42E-08 & {\bf 0.0\%} \\ 
        7 & $\pm25\%$ & 1.05E-06 & {\bf 0.0\%} & 2.46E-08 & {\bf 0.0\%} & 9.14E-07 & {\bf 0.0\%} & 2.23E-08 & {\bf 0.0\%} & 3.60E-07 & {\bf 0.0\%} & 2.89E-09 & {\bf 0.0\%} \\ 
        8 & $\pm25\%$ & 1.15E-07 & {\bf 0.0\%} & 1.21E-09 & {\bf 0.0\%} & 1.02E-07 & {\bf 0.0\%} & 1.12E-09 & {\bf 0.0\%} & 3.49E-08 & {\bf 0.0\%} & 1.52E-10 & {\bf 0.0\%} \\ 
        9 & $\pm25\%$ & 1.15E-08 & {\bf 0.0\%} & 5.32E-11 & {\bf 0.0\%} & 1.05E-08 & {\bf 0.0\%} & 5.05E-11 & {\bf 0.0\%} & 3.12E-09 & {\bf 0.0\%} & 8.26E-12 & {\bf 0.0\%} \\ 
        10 & $\pm25\%$ & 1.08E-09 & {\bf 0.0\%} & 6.76E-12 & {\bf 0.0\%} & 1.01E-09 & {\bf 0.0\%} & 6.07E-12 & {\bf 0.0\%} & 2.57E-10 & {\bf 0.0\%} & 4.96E-13 & {\bf 0.0\%} \\ \hline
        1 & $\pm50\%$ & 3.51E-01 & 35.1\% & 1.13E-01 & 11.3\% & 3.42E-01 & 34.2\% & 1.07E-01 & 10.7\% & 3.41E-01 & 34.1\% & 1.04E-01 & 10.4\% \\ 
        2 & $\pm50\%$ & 5.27E-01 & 52.7\% & 8.21E-01 & 82.1\% & 5.23E-01 & 52.3\% & 7.08E-01 & 70.8\% & 5.31E-01 & 53.1\% & 7.84E-01 & 78.4\% \\ 
        3 & $\pm50\%$ & 7.28E-02 & 7.3\% & 9.29E-03 & 0.9\% & 6.68E-02 & 6.7\% & 8.41E-03 & 0.8\% & 6.58E-02 & 6.6\% & 5.36E-03 & 0.5\% \\ 
        4 & $\pm50\%$ & 2.06E-01 & 20.6\% & 1.71E-03 & 0.2\% & 2.05E-01 & 20.5\% & 1.83E-03 & 0.2\% & 2.05E-01 & 20.5\% & 1.12E-03 & 0.1\% \\
        5 & $\pm50\%$ & 4.50E-03 & 0.5\% & 2.88E-04 & {\bf 0.0\%} & 3.44E-03 & 0.3\% & 3.14E-04 & {\bf 0.0\%} & 2.48E-03 & 0.2\% & 1.20E-04 & {\bf 0.0\%} \\ 
        6 & $\pm50\%$ & 3.14E-02 & 3.1\% & 3.83E-05 & {\bf 0.0\%} & 3.10E-02 & 3.1\% & 4.28E-05 & {\bf 0.0\%} & 3.06E-02 & 3.1\% & 1.43E-05 & {\bf 0.0\%} \\ 
        7 & $\pm50\%$ & 3.53E-04 & {\bf 0.0\%} & 5.81E-06 & {\bf 0.0\%} & 2.67E-04 & {\bf 0.0\%} & 4.52E-06 & {\bf 0.0\%} & 1.60E-04 & {\bf 0.0\%} & 1.50E-06 & {\bf 0.0\%} \\ 
        8 & $\pm50\%$ & 7.69E-05 & {\bf 0.0\%} & 6.41E-07 & {\bf 0.0\%} & 5.89E-05 & {\bf 0.0\%} & 5.14E-07 & {\bf 0.0\%} & 3.11E-05 & {\bf 0.0\%} & 1.40E-07 & {\bf 0.0\%} \\
        9 & $\pm50\%$ & 1.41E-05 & {\bf 0.0\%} & 6.35E-08 & {\bf 0.0\%} & 1.12E-05 & {\bf 0.0\%} & 5.41E-08 & {\bf 0.0\%} & 5.24E-06 & {\bf 0.0\%} & 1.28E-08 & {\bf 0.0\%} \\ 
        10 & $\pm50\%$ & 2.43E-06 & {\bf 0.0\%} & 5.63E-09 & {\bf 0.0\%} & 2.00E-06 & {\bf 0.0\%} & 4.58E-09 & {\bf 0.0\%} & 8.16E-07 & {\bf 0.0\%} & 1.04E-09 & {\bf 0.0\%} \\ \hline
    \end{tabular}
    \caption{Numerical results. The first column $N$ is the degree of the Taylor polynomial calculated. In column 2, $\pm\%$ relatively scales fixed parameter values to create intervals where test parameters are randomly selected, e.g., $\pm5\% \equiv [0.95x\quad1.05x]$. The remaining columns are the raw and percentage values of relative error between the actual parameters and those found by the algorithm. The instances of `0.0\%' error represent the algorithm finding parameter values with high accuracy and are bolded.}
    \label{tab:numericalresults1}
\end{table}
    
\newpage
\begin{table}[!ht]
    \centering
    \tiny
    \begin{tabular}{|l|l|l|l|l|l|l|l|l|l|l|l|l|l|}
    \hline
        $N$ & Interval & Mean $\mu_L$ & & Med $\mu_L$ & & Mean $c_{PA}$ &  & Med $c_{PA}$ &  & Mean $\mu_A$&  & Med $\mu_A$ &  \\ \hline \hline
        1 & $\pm5\%$ & 5.23E-16 & {\bf 0.0\%} & 5.27E-16 & {\bf 0.0\%} & 1.00E+00 & 100.0\% & 1.00E+00 & 100.0\% & 1.00E+00 & 100.0\% & 1.00E+00 & 100.0\% \\ 
        2 & $\pm5\%$ & 5.26E-16 & {\bf 0.0\%} & 5.24E-16 & {\bf 0.0\%} & 1.00E+00 & 100.0\% & 1.00E+00 & 100.0\% & 1.00E+00 & 100.0\% & 1.00E+00 & 100.0\% \\ 
        3 & $\pm5\%$ & 5.27E-16 & {\bf 0.0\%} & 5.24E-16 & {\bf 0.0\%} & 7.23E-02 & 7.2\% & 2.47E-03 & 0.2\% & 7.49E-02 & 7.5\% & 5.46E-03 & 0.5\% \\ 
        4 & $\pm5\%$ & 5.26E-16 & {\bf 0.0\%} & 5.27E-16 & {\bf 0.0\%} & 1.02E-02 & 1.0\% & 2.41E-04 & {\bf 0.0\%} & 1.14E-02 & 1.1\% & 1.38E-03 & 0.1\% \\
        5 & $\pm5\%$ & 5.27E-16 & {\bf 0.0\%} & 5.24E-16 & {\bf 0.0\%} & 1.64E-05 & {\bf 0.0\%} & 1.61E-05 & {\bf 0.0\%} & 3.77E-05 & {\bf 0.0\%} & 3.76E-05 & {\bf 0.0\%} \\ 
        6 & $\pm5\%$ & 5.24E-16 & {\bf 0.0\%} & 5.24E-16 & {\bf 0.0\%} & 1.18E-06 & {\bf 0.0\%} & 1.14E-06 & {\bf 0.0\%} & 6.78E-06 & {\bf 0.0\%} & 6.76E-06 & {\bf 0.0\%} \\ 
        7 & $\pm5\%$ & 5.25E-16 & {\bf 0.0\%} & 5.24E-16 & {\bf 0.0\%} & 5.98E-08 & {\bf 0.0\%} & 5.75E-08 & {\bf 0.0\%} & 1.45E-07 & {\bf 0.0\%} & 1.40E-07 & {\bf 0.0\%} \\ 
        8 & $\pm5\%$ & 5.27E-16 & {\bf 0.0\%} & 5.24E-16 & {\bf 0.0\%} & 3.36E-09 & {\bf 0.0\%} & 3.20E-09 & {\bf 0.0\%} & 1.94E-08 & {\bf 0.0\%} & 1.93E-08 & {\bf 0.0\%} \\ 
        9 & $\pm5\%$ & 5.26E-16 & {\bf 0.0\%} & 5.24E-16 & {\bf 0.0\%} & 1.37E-10 & {\bf 0.0\%} & 1.29E-10 & {\bf 0.0\%} & 3.48E-10 & {\bf 0.0\%} & 3.20E-10 & {\bf 0.0\%} \\ 
        10 & $\pm5\%$ & 5.29E-16 & {\bf 0.0\%} & 5.27E-16 & {\bf 0.0\%} & 6.32E-12 & {\bf 0.0\%} & 5.89E-12 & {\bf 0.0\%} & 3.63E-11 & {\bf 0.0\%} & 3.59E-11 & {\bf 0.0\%} \\ \hline
        1 & $\pm10\%$ & 4.68E-16 & {\bf 0.0\%} & 4.08E-16 & {\bf 0.0\%} & 9.92E-01 & 99.2\% & 1.00E+00 & 100.0\% & 9.91E-01 & 99.1\% & 1.00E+00 & 100.0\% \\ 
        2 & $\pm10\%$ & 4.69E-16 & {\bf 0.0\%} & 4.08E-16 & {\bf 0.0\%} & 6.59E-01 & 65.9\% & 1.00E+00 & 100.0\% & 7.00E-01 & 7{\bf 0.0\%} & 1.00E+00 & 100.0\% \\ 
        3 & $\pm10\%$ & 4.69E-16 & {\bf 0.0\%} & 4.08E-16 & {\bf 0.0\%} & 6.24E-02 & 6.2\% & 2.54E-03 & 0.3\% & 6.50E-02 & 6.5\% & 5.56E-03 & 0.6\% \\ 
        4 & $\pm10\%$ & 4.67E-16 & {\bf 0.0\%} & 4.08E-16 & {\bf 0.0\%} & 1.03E-02 & 1.0\% & 2.48E-04 & {\bf 0.0\%} & 1.14E-02 & 1.1\% & 1.42E-03 & 0.1\% \\ 
        5 & $\pm10\%$ & 4.69E-16 & {\bf 0.0\%} & 4.08E-16 & {\bf 0.0\%} & 1.78E-05 & {\bf 0.0\%} & 1.67E-05 & {\bf 0.0\%} & 4.49E-05 & {\bf 0.0\%} & 3.90E-05 & {\bf 0.0\%} \\ 
        6 & $\pm10\%$ & 4.67E-16 & {\bf 0.0\%} & 4.08E-16 & {\bf 0.0\%} & 1.30E-06 & {\bf 0.0\%} & 1.19E-06 & {\bf 0.0\%} & 7.23E-06 & {\bf 0.0\%} & 7.09E-06 & {\bf 0.0\%} \\ 
        7 & $\pm10\%$ & 4.69E-16 & {\bf 0.0\%} & 4.08E-16 & {\bf 0.0\%} & 6.83E-08 & {\bf 0.0\%} & 6.05E-08 & {\bf 0.0\%} & 2.01E-07 & {\bf 0.0\%} & 1.67E-07 & {\bf 0.0\%} \\ 
        8 & $\pm10\%$ & 4.67E-16 & {\bf 0.0\%} & 4.08E-16 & {\bf 0.0\%} & 3.92E-09 & {\bf 0.0\%} & 3.37E-09 & {\bf 0.0\%} & 2.15E-08 & {\bf 0.0\%} & 2.06E-08 & {\bf 0.0\%} \\ 
        9 & $\pm10\%$ & 4.67E-16 & {\bf 0.0\%} & 4.08E-16 & {\bf 0.0\%} & 1.67E-10 & {\bf 0.0\%} & 1.37E-10 & {\bf 0.0\%} & 5.51E-10 & {\bf 0.0\%} & 4.74E-10 & {\bf 0.0\%} \\ 
        10 & $\pm10\%$ & 4.69E-16 & {\bf 0.0\%} & 4.08E-16 & {\bf 0.0\%} & 7.87E-12 & {\bf 0.0\%} & 6.29E-12 & {\bf 0.0\%} & 4.26E-11 & {\bf 0.0\%} & 3.98E-11 & {\bf 0.0\%} \\ \hline
        1 & $\pm20\%$ & 4.94E-16 & {\bf 0.0\%} & 4.81E-16 & {\bf 0.0\%} & 8.52E-01 & 85.2\% & 1.00E+00 & 100.0\% & 8.38E-01 & 83.8\% & 1.00E+00 & 100.0\% \\
        2 & $\pm20\%$ & 4.94E-16 & {\bf 0.0\%} & 4.81E-16 & {\bf 0.0\%} & 3.09E-01 & 30.9\% & 3.21E-02 & 3.2\% & 3.95E-01 & 39.5\% & 1.64E-01 & 16.4\% \\ 
        3 & $\pm20\%$ & 4.94E-16 & {\bf 0.0\%} & 4.81E-16 & {\bf 0.0\%} & 4.28E-02 & 4.3\% & 2.77E-03 & 0.3\% & 4.72E-02 & 4.7\% & 6.17E-03 & 0.6\% \\ 
        4 & $\pm20\%$ & 4.96E-16 & {\bf 0.0\%} & 4.81E-16 & {\bf 0.0\%} & 3.01E-04 & {\bf 0.0\%} & 2.45E-04 & {\bf 0.0\%} & 1.61E-03 & 0.2\% & 1.50E-03 & 0.1\% \\ 
        5 & $\pm20\%$ & 4.96E-16 & {\bf 0.0\%} & 4.81E-16 & {\bf 0.0\%} & 2.23E-05 & {\bf 0.0\%} & 1.65E-05 & {\bf 0.0\%} & 7.81E-05 & {\bf 0.0\%} & 5.75E-05 & {\bf 0.0\%} \\ 
        6 & $\pm20\%$ & 4.97E-16 & {\bf 0.0\%} & 4.81E-16 & {\bf 0.0\%} & 1.73E-06 & {\bf 0.0\%} & 1.17E-06 & {\bf 0.0\%} & 9.08E-06 & {\bf 0.0\%} & 8.04E-06 & {\bf 0.0\%} \\ 
        7 & $\pm20\%$ & 4.96E-16 & {\bf 0.0\%} & 4.81E-16 & {\bf 0.0\%} & 1.00E-07 & {\bf 0.0\%} & 5.95E-08 & {\bf 0.0\%} & 3.98E-07 & {\bf 0.0\%} & 2.84E-07 & {\bf 0.0\%} \\ 
        8 & $\pm20\%$ & 4.96E-16 & {\bf 0.0\%} & 4.81E-16 & {\bf 0.0\%} & 6.15E-09 & {\bf 0.0\%} & 3.30E-09 & {\bf 0.0\%} & 3.11E-08 & {\bf 0.0\%} & 2.59E-08 & {\bf 0.0\%} \\ 
        9 & $\pm20\%$ & 4.94E-16 & {\bf 0.0\%} & 4.81E-16 & {\bf 0.0\%} & 2.97E-10 & {\bf 0.0\%} & 1.34E-10 & {\bf 0.0\%} & 1.24E-09 & {\bf 0.0\%} & 8.58E-10 & {\bf 0.0\%} \\ 
        10 & $\pm20\%$ & 4.96E-16 & {\bf 0.0\%} & 4.81E-16 & {\bf 0.0\%} & 1.50E-11 & {\bf 0.0\%} & 6.09E-12 & {\bf 0.0\%} & 7.21E-11 & {\bf 0.0\%} & 5.49E-11 & {\bf 0.0\%} \\ \hline
        1 & $\pm25\%$ & 5.10E-16 & {\bf 0.0\%} & 4.62E-16 & {\bf 0.0\%} & 7.64E-01 & 76.4\% & 1.00E+00 & 100.0\% & 7.47E-01 & 74.7\% & 1.00E+00 & 100.0\% \\ 
        2 & $\pm25\%$ & 5.12E-16 & {\bf 0.0\%} & 4.71E-16 & {\bf 0.0\%} & 4.08E-01 & 40.8\% & 4.33E-02 & 4.3\% & 4.81E-01 & 48.1\% & 1.76E-01 & 17.6\% \\ 
        3 & $\pm25\%$ & 5.11E-16 & {\bf 0.0\%} & 4.62E-16 & {\bf 0.0\%} & 4.33E-02 & 4.3\% & 2.71E-03 & 0.3\% & 4.80E-02 & 4.8\% & 7.64E-03 & 0.8\% \\ 
        4 & $\pm25\%$ & 5.14E-16 & {\bf 0.0\%} & 4.62E-16 & {\bf 0.0\%} & 1.04E-02 & 1.0\% & 2.58E-04 & {\bf 0.0\%} & 1.17E-02 & 1.2\% & 1.55E-03 & 0.2\% \\ 
        5 & $\pm25\%$ & 5.14E-16 & {\bf 0.0\%} & 4.71E-16 & {\bf 0.0\%} & 2.85E-05 & {\bf 0.0\%} & 1.76E-05 & {\bf 0.0\%} & 9.17E-05 & {\bf 0.0\%} & 7.33E-05 & {\bf 0.0\%} \\ 
        6 & $\pm25\%$ & 5.14E-16 & {\bf 0.0\%} & 4.71E-16 & {\bf 0.0\%} & 2.27E-06 & {\bf 0.0\%} & 1.26E-06 & {\bf 0.0\%} & 1.00E-05 & {\bf 0.0\%} & 8.94E-06 & {\bf 0.0\%} \\ 
        7 & $\pm25\%$ & 5.12E-16 & {\bf 0.0\%} & 4.62E-16 & {\bf 0.0\%} & 1.41E-07 & {\bf 0.0\%} & 6.49E-08 & {\bf 0.0\%} & 4.93E-07 & {\bf 0.0\%} & 3.92E-07 & {\bf 0.0\%} \\ 
        8 & $\pm25\%$ & 5.11E-16 & {\bf 0.0\%} & 4.62E-16 & {\bf 0.0\%} & 8.87E-09 & {\bf 0.0\%} & 3.61E-09 & {\bf 0.0\%} & 3.67E-08 & {\bf 0.0\%} & 3.09E-08 & {\bf 0.0\%} \\ 
        9 & $\pm25\%$ & 5.10E-16 & {\bf 0.0\%} & 4.62E-16 & {\bf 0.0\%} & 4.62E-10 & {\bf 0.0\%} & 1.50E-10 & {\bf 0.0\%} & 1.63E-09 & {\bf 0.0\%} & 1.22E-09 & {\bf 0.0\%} \\ 
        10 & $\pm25\%$ & 5.08E-16 & {\bf 0.0\%} & 4.62E-16 & {\bf 0.0\%} & 2.38E-11 & {\bf 0.0\%} & 6.80E-12 & {\bf 0.0\%} & 9.20E-11 & {\bf 0.0\%} & 6.92E-11 & {\bf 0.0\%} \\ \hline
        1 & $\pm50\%$ & 5.99E-16 & {\bf 0.0\%} & 5.68E-16 & {\bf 0.0\%} & 6.54E-01 & 65.4\% & 1.00E+00 & 100.0\% & 7.28E-01 & 72.8\% & 1.00E+00 & 100.0\% \\ 
        2 & $\pm50\%$ & 5.96E-16 & {\bf 0.0\%} & 5.62E-16 & {\bf 0.0\%} & 3.18E-01 & 31.8\% & 5.59E-02 & 5.6\% & 4.30E-01 & 43.0\% & 2.28E-01 & 22.8\% \\ 
        3 & $\pm50\%$ & 5.96E-16 & {\bf 0.0\%} & 5.62E-16 & {\bf 0.0\%} & 2.57E-02 & 2.6\% & 2.36E-03 & 0.2\% & 3.96E-02 & 4.0\% & 1.91E-02 & 1.9\% \\ 
        4 & $\pm50\%$ & 5.96E-16 & {\bf 0.0\%} & 5.62E-16 & {\bf 0.0\%} & 7.65E-04 & 0.1\% & 2.00E-04 & {\bf 0.0\%} & 3.08E-03 & 0.3\% & 2.70E-03 & 0.3\% \\ 
        5 & $\pm50\%$ & 6.02E-16 & {\bf 0.0\%} & 5.68E-16 & {\bf 0.0\%} & 8.00E-05 & {\bf 0.0\%} & 1.30E-05 & {\bf 0.0\%} & 2.84E-04 & {\bf 0.0\%} & 2.31E-04 & {\bf 0.0\%} \\ 
        6 & $\pm50\%$ & 5.99E-16 & {\bf 0.0\%} & 5.68E-16 & {\bf 0.0\%} & 7.69E-06 & {\bf 0.0\%} & 9.11E-07 & {\bf 0.0\%} & 2.76E-05 & {\bf 0.0\%} & 2.10E-05 & {\bf 0.0\%} \\ 
        7 & $\pm50\%$ & 6.00E-16 & {\bf 0.0\%} & 5.63E-16 & {\bf 0.0\%} & 6.42E-07 & {\bf 0.0\%} & 6.99E-08 & {\bf 0.0\%} & 2.10E-06 & {\bf 0.0\%} & 1.38E-06 & {\bf 0.0\%} \\ 
        8 & $\pm50\%$ & 5.99E-16 & {\bf 0.0\%} & 5.63E-16 & {\bf 0.0\%} & 4.92E-08 & {\bf 0.0\%} & 4.39E-09 & {\bf 0.0\%} & 1.55E-07 & {\bf 0.0\%} & 9.08E-08 & {\bf 0.0\%} \\ 
        9 & $\pm50\%$ & 6.02E-16 & {\bf 0.0\%} & 5.68E-16 & {\bf 0.0\%} & 1.00E-02 & 1.0\% & 2.32E-10 & {\bf 0.0\%} & 1.00E-02 & 1.0\% & 4.85E-09 & {\bf 0.0\%} \\ 
        10 & $\pm50\%$ & 5.98E-16 & {\bf 0.0\%} & 5.62E-16 & {\bf 0.0\%} & 1.00E-02 & 1.0\% & 1.06E-11 & {\bf 0.0\%} & 1.00E-02 & 1.0\% & 2.55E-10 & {\bf 0.0\%} \\ \hline
    \end{tabular}
    \caption{Numerical results. The first column $N$ is the degree of the Taylor polynomial calculated. In column 2, $\pm\%$ relatively scales fixed parameter values to create intervals where test parameters are randomly selected, e.g., $\pm5\% \equiv [0.95x\quad1.05x]$. The remaining columns are the raw and percentage values of relative error between the actual parameters and those found by the algorithm. Note that since the algorithm sometimes returns 0 (no solution found), the relative error appears as `100\%'. The instances of `0.0\%' error represent the algorithm finding parameter values with high accuracy and are bolded.}
    \label{tab:numericalresults2}
\end{table}
\newpage

\subsubsection{Applied to competition models}

We applied our method to the competition model as in equation ~\eqref{eq:DLVM}, i.e. the discrete Lotka-Volterra model. The experimental set up is much the same as our approach in the LPA model: pre-selected values of the parameters, synthesized data, and increasing length intervals of parameter `guesses' as would be input by the user (endpoints perturbed to create non-symmetric intervals). Provided this data we then run our algorithms. The results are in Tables~\ref{tab:compmodeldim2mean}, \ref{tab:compmodeldim2median}, \ref{tab:compmodeldim3mean}, and \ref{tab:compmodeldim3median}, presented as mean and median relative error. 

The `dimension' of the competition model is the value $n$ where $i=1,\ldots,n$. We ran the experiment on dimensions 2 and 3 models. This experiment has available Julia implementation as linked in the introduction to Section~\ref{sect:experiment}. Readers following along should note that in the code, the dimension is labeled $n$, with degree of Taylor expansion as $q$. The paper by comparison follows its own previous notation: in each Table~\ref{tab:compmodeldim2mean}, \ref{tab:compmodeldim2median}, \ref{tab:compmodeldim3mean}, and \ref{tab:compmodeldim3median}, $n$ represents the number of terms in the Taylor expansion.

\subsubsection{Dimension 2}
In the dimension 2 case the experiment was run with 100 trials and up to 7 terms of the Taylor series, labeled as the first columns in tables ~\ref{tab:compmodeldim2mean} and ~\ref{tab:compmodeldim2median}. Our algorithm almost always converges to 4 correct digits within 3 Taylor series terms, and in the exception reached this within 5 terms. Our algorithm was correct to within 10 digits in most cases by 6 terms, often fewer.

\begin{table}[!ht]

\centering
\footnotesize
\begin{tabular}{|l|l|l|l|l|l|l|l|} 
\hline
$N$        & Interval      & Mean $a_{11}$       & Mean $a_{12}$       & Mean $a_{21}$       & Mean $a_{22}$       & Mean $r_1$        & Mean $r_2$         \\ 
\hline
\hline
1 & 0.05 & 1.0E-01 & 1.0E-01 & 1.0E-01 & 2.9E-04 & 6.4E-04 & 3.1E-04\\
\bf 2 & \bf 0.05 &\bf  1.8E-06 &\bf  9.1E-06 &\bf  2.1E-06 &\bf  3.3E-06 &\bf  1.5E-05 & \bf 6.9E-06\\
\bf 3 &\bf  0.05 &\bf  1.3E-08 &\bf  8.7E-08 &\bf  2.2E-08 &\bf  2.6E-08 &\bf  1.7E-07 &\bf  7.9E-08\\
\bf 4 &\bf  0.05 &\bf  8.3E-11 &\bf  6.5E-10 &\bf  1.8E-10 &\bf  1.6E-10 &\bf  1.3E-09 &\bf  6.2E-10\\
\bf 5 &\bf  0.05 &\bf  4.4E-13 &\bf  4.1E-12 &\bf  1.2E-12 &\bf  7.6E-13 &\bf  7.5E-12 &\bf  3.7E-12\\
\hline
1 & 0.1 & 2.0E-01 & 2.1E-01 & 2.0E-01 & 1.0E-01 & 1.0E-01 & 1.0E-01\\
 2 & 0.1 &  5.0E-05 & 5.7E-04 & 3.2E-04 & 1.5E-04 &2.0E-03 & 1.6E-03\\
\bf 3 &\bf  0.1 &\bf  7.7E-07 &\bf  1.2E-05 &\bf  6.8E-06 &\bf  3.9E-06 &\bf  6.2E-05 &\bf  4.8E-05\\
\bf 4 &\bf  0.1 &\bf  9.6E-09 &\bf  1.8E-07 &\bf  1.0E-07 &\bf  7.8E-08 &\bf  1.4E-06 &\bf  1.1E-06\\
\bf 5 &\bf  0.1 &\bf  1.1E-10 &\bf  2.2E-09 &\bf  1.2E-09 &\bf  1.3E-09 &\bf  2.5E-08 &\bf  2.0E-08\\
\hline
1 & 0.2 & 3.6E-03 & 1.4E-02 & 1.1E-02 & 1.1E-01 & 1.1E-01 & 1.1E-01\\
2 &0.2 & 8.7E-05 & 5.8E-04 & 4.3E-04 & 5.0E-04 & 8.5E-04 & 1.0E-03\\
\bf 3 &\bf  0.2 &\bf  1.5E-06 & \bf 1.1E-05 &\bf  8.4E-06 &\bf  2.2E-05 &\bf  5.3E-05 &\bf  6.3E-05\\
\bf 4 &\bf  0.2 &\bf  2.1E-08 &\bf  1.6E-07 & \bf 1.2E-07 &\bf  8.6E-07 &\bf  2.6E-06 &\bf  3.3E-06\\
\bf 5 &\bf  0.2 &\bf  2.7E-10 &\bf  1.7E-09 &\bf  1.4E-09 &\bf  2.8E-08 &\bf  1.0E-07 &\bf  1.4E-07\\
\hline
1 & 0.25 & 1.0E-01 & 1.1E-01 & 1.1E-01 & 2.1E-01 & 2.2E-01 & 2.2E-01\\
2 & 0.25 & 1.0E-01 & 1.0E-01 & 1.0E-01 & 2.0E-01 & 2.0E-01 & 2.0E-01\\
 3 & 0.25 & 8.3E-05 & 4.7E-04 &  2.3E-04 & 4.8E-04 &3.1E-03 &2.3E-03\\
\bf 4 &\bf  0.25 &\bf  2.0E-06 &\bf  1.1E-05 &\bf  5.5E-06 &\bf  2.0E-05 &\em 1.3E-04 &\bf  9.3E-05\\
\bf 5 &\bf  0.25 &\bf  3.6E-08 &\bf  2.1E-07 &\bf  9.9E-08 &\bf  6.4E-07 &\bf  4.2E-06 &\bf  3.0E-06\\
\hline
1 & 0.5 & 1.4E-01 & 1.5E-01 & 1.2E-01 & 3.4E-01 & 3.5E-01 & 3.3E-01\\
2 & 0.5 & 8.8E-03 & 4.2E-03 & 2.9E-03 & 2.3E-02 & 1.2E-02 & 1.2E-02\\
3 & 0.5 & 7.0E-04 & 2.7E-04 & 2.1E-04 & 2.4E-03 & 1.2E-03 & 1.2E-03\\
\bf 4 &\bf  0.5 &\bf  4.5E-05 &\bf  1.4E-05 &\bf  1.4E-05 &\em  2.0E-04 &\bf  9.1E-05 &\bf  9.0E-05\\
\bf 5 &\bf  0.5 &\bf  2.2E-06 &\bf  6.3E-07 &\bf  6.9E-07 &\bf  1.3E-05 &\bf  5.9E-06 &\bf  5.8E-06\\

\hline
\end{tabular}\caption{Numerical data representing {\bf mean} relative error in the approximated system to the actual parameter values in a 2-dimensional discrete Lotka-Volterra model. The first column $N$ is the degree (number of terms) of the Taylor expansion. Rows are bolded whenever the algorithm found all parameter values within 6 significant digits, mean.}
    \label{tab:compmodeldim2mean}
\end{table}
\newpage
\begin{table}[!ht]
\centering
\footnotesize
\begin{tabular}{|l|l|l|l|l|l|l|l|} 
\hline
$N$        & Interval      & Med $a_{11}$        & Med $a_{12}$        & Med $a_{21}$        & Med $a_{22}$        & Med $r_1$         & Med $r_2$          \\ 
\hline
\hline
1 & 0.05 & 1.0E-01 & 1.0E-01 & 1.0E-01 & 1.0E-01 & 2.7E-04 & 1.0E-01 \\
\bf 2 &\bf  0.05 &\bf   1.8E-05 &\em  1.0E-04 &\bf  3.0E-05 &\bf  2.0E-06 &\bf  3.5E-06 &\bf  3.6E-05 \\
\bf 3 &\bf  0.05 &\bf  2.8E-07 &\bf  2.0E-06 &\bf  6.0E-07 &\bf  1.7E-08 &\bf  3.9E-08 &\bf  7.9E-07 \\
\bf 4 &\bf  0.05 &\bf  3.5E-09 &\bf  3.2E-08 &\bf  9.6E-09 &\bf  1.3E-10 &\bf  3.1E-10 &\bf  1.3E-08 \\
\bf 5 &\bf  0.05 &\bf  3.8E-11 &\bf  4.4E-10 &\bf  1.3E-10 &\bf  8.0E-13 &\bf  1.9E-12 &\bf  1.8E-10 \\
\hline
1 & 0.1 &  2.0E-01 & 2.1E-01 & 2.1E-01 & 2.0E-01 & 1.0E-01 & 2.0E-01\\
2 & 0.1 &  2.0E-01 & 2.0E-01 & 2.0E-01 & 5.5E-05 & 2.2E-04 & 2.0E-01 \\
\bf 3 &\bf  0.1 &\bf   2.3E-05 &\em  4.9E-04 &\bf  3.5E-04 &\bf  1.3E-06 &\bf  6.7E-06 &\bf  6.0E-05 \\
\bf 4 &\bf  0.1 &\em 1.0E-01 &\em 1.0E-01 &\em 1.0E-01 &\bf  2.1E-08 &\bf  1.4E-07 &\em 1.0E-01\\
\bf 5 &\bf  0.1 &\bf   1.7E-08 &\bf  3.6E-07 &\bf  2.6E-07 &\bf  2.6E-10 &\bf  2.6E-09 &\bf  4.8E-08\\
\hline
1 & 0.2 &  2.1E-01 & 2.1E-01 & 2.1E-01 & 2.1E-03 & 1.0E-01 & 2.0E-01\\
 2 & 0.2 &  3.3E-04 & 1.5E-03 & 1.3E-03 &  4.7E-05 & 3.4E-04 &  2.7E-04\\
\bf 3 & 0.2 &\bf   1.2E-05 &\bf  4.5E-05 &\bf  3.8E-05 &\bf  8.8E-07 &\bf  2.1E-05 &\bf  7.5E-06\\
\bf 4 & 0.2 &\bf   3.4E-07 &\bf  1.2E-06 &\bf  8.5E-07 &\bf  1.3E-08 &\bf  9.4E-07 &\bf  2.0E-07\\
\bf 5 & 0.2 &\bf   8.4E-09 &\bf  2.6E-08 &\bf  1.7E-08 &\bf  1.5E-10 &\bf  3.3E-08 &\bf  4.3E-09\\
\hline
1 & 0.25 & 1.1E-01 & 1.3E-01 & 1.1E-01 & 1.0E-01 & 2.0E-01 & 1.1E-01\\
2 & 0.25 & 2.0E-01 & 2.0E-01 & 2.0E-01 & 1.0E-01 & 2.0E-01 & 2.0E-01\\
 3 &  0.25 &  1.8E-04 & 1.5E-03 &  8.2E-04 &  2.1E-05 &  3.2E-04 &  4.2E-04\\
\bf 4 &\bf  0.25 &\bf  1.0E-05 &\bf  7.2E-05 &\bf  3.6E-05 &\bf  4.9E-07 &\bf  1.5E-05 &\bf  3.2E-05\\
\bf 5 &\bf  0.25 &\bf   5.7E-07 &\bf  3.5E-06 &\bf  1.6E-06 &\bf  8.7E-09 &\bf  5.3E-07 &\bf  2.1E-06\\
\hline
1 & 0.5 &  3.4E-01 & 3.6E-01 & 3.6E-01 & 1.2E-01 & 3.3E-01 & 3.4E-01\\
2 & 0.5 &  3.1E-01 & 3.1E-01 & 3.1E-01 & 2.1E-03 & 8.3E-03 & 3.1E-01\\
3 & 0.5 & 2.0E-01 & 2.0E-01 & 2.0E-01 & 1.5E-04 & 7.9E-04 & 2.0E-01\\
\bf 4 &\bf  0.5 &\em  1.0E-03 &\em 3.2E-03 &\em 3.4E-03 &\bf  9.1E-06 &\bf  6.4E-05 &\em 2.3E-03\\
\bf  5 &\bf  0.5 &\em   1.4E-04 &\em  4.6E-04 &\em  4.7E-04 &\bf   4.6E-07 &\bf   4.3E-06 &\em  3.3E-04\\
\hline
\end{tabular}\caption{Numerical data representing {\bf median} relative error in the approximated system to the actual parameter values in a 2-dimensional discrete Lotka-Volterra model. The first column $N$ is the degree (number of terms) of the Taylor expansion. Rows are bolded whenever the algorithm found all parameter values within 6 significant digits, median.}
    \label{tab:compmodeldim2median}
\end{table}

\subsubsection{Dimension 3}

In dimension 3, the increase in number of parameters increased computation time and so we scaled back. In this case 10 instances were run on each round, up to 5 terms in the Taylor series. In nearly all cases our approach was correct to 5 digits within 5 degrees of Taylor expansion.
\newpage
\begin{table}[!ht]
\scriptsize
\centering
\begin{tabular}{|l|l|l|l|l|l|l|l|} 
\hline
$N$        & Interval      & Mean $a_{11}$       & Mean $a_{12}$       & Mean $a_{13}$       & Mean $a_{21}$          & Mean $a_{22}$          & Mean $a_{23}$            \\ 
\hline
\hline
1 & 0.05 & 1.0E-01 & 1.0E-01 & 1.0E-01 & 2.9E-04 & 6.4E-04 & 3.1E-04\\
\bf 2 & \bf 0.05 &\bf  1.8E-06 &\bf  9.1E-06 &\bf  2.1E-06 &\bf  3.3E-06 &\bf  1.5E-05 & \bf 6.9E-06\\
\bf 3 &\bf  0.05 &\bf  1.3E-08 &\bf  8.7E-08 &\bf  2.2E-08 &\bf  2.6E-08 &\bf  1.7E-07 &\bf  7.9E-08\\
\bf 4 &\bf  0.05 &\bf  8.3E-11 &\bf  6.5E-10 &\bf  1.8E-10 &\bf  1.6E-10 &\bf  1.3E-09 &\bf  6.2E-10\\
\bf 5 &\bf  0.05 &\bf  4.4E-13 &\bf  4.1E-12 &\bf  1.2E-12 &\bf  7.6E-13 &\bf  7.5E-12 &\bf  3.7E-12\\
\hline
1 & 0.1 & 2.0E-01 & 2.1E-01 & 2.0E-01 & 1.0E-01 & 1.0E-01 & 1.0E-01\\
 2 & 0.1 &  5.0E-05 & 5.7E-04 & 3.2E-04 & 1.5E-04 &2.0E-03 & 1.6E-03\\
\bf 3 &\bf  0.1 &\bf  7.7E-07 &\bf  1.2E-05 &\bf  6.8E-06 &\bf  3.9E-06 &\bf  6.2E-05 &\bf  4.8E-05\\
\bf 4 &\bf  0.1 &\bf  9.6E-09 &\bf  1.8E-07 &\bf  1.0E-07 &\bf  7.8E-08 &\bf  1.4E-06 &\bf  1.1E-06\\
\bf 5 &\bf  0.1 &\bf  1.1E-10 &\bf  2.2E-09 &\bf  1.2E-09 &\bf  1.3E-09 &\bf  2.5E-08 &\bf  2.0E-08\\
\hline
1 & 0.2 & 3.6E-03 & 1.4E-02 & 1.1E-02 & 1.1E-01 & 1.1E-01 & 1.1E-01\\
2 &0.2 & 8.7E-05 & 5.8E-04 & 4.3E-04 & 5.0E-04 & 8.5E-04 & 1.0E-03\\
\bf 3 &\bf  0.2 &\bf  1.5E-06 & \bf 1.1E-05 &\bf  8.4E-06 &\bf  2.2E-05 &\bf  5.3E-05 &\bf  6.3E-05\\
\bf 4 &\bf  0.2 &\bf  2.1E-08 &\bf  1.6E-07 & \bf 1.2E-07 &\bf  8.6E-07 &\bf  2.6E-06 &\bf  3.3E-06\\
\bf 5 &\bf  0.2 &\bf  2.7E-10 &\bf  1.7E-09 &\bf  1.4E-09 &\bf  2.8E-08 &\bf  1.0E-07 &\bf  1.4E-07\\
\hline
1 & 0.25 & 1.0E-01 & 1.1E-01 & 1.1E-01 & 2.1E-01 & 2.2E-01 & 2.2E-01\\
2 & 0.25 & 1.0E-01 & 1.0E-01 & 1.0E-01 & 2.0E-01 & 2.0E-01 & 2.0E-01\\
 3 & 0.25 & 8.3E-05 & 4.7E-04 &  2.3E-04 & 4.8E-04 &3.1E-03 &2.3E-03\\
\bf 4 &\bf  0.25 &\bf  2.0E-06 &\bf  1.1E-05 &\bf  5.5E-06 &\bf  2.0E-05 &\em 1.3E-04 &\bf  9.3E-05\\
\bf 5 &\bf  0.25 &\bf  3.6E-08 &\bf  2.1E-07 &\bf  9.9E-08 &\bf  6.4E-07 &\bf  4.2E-06 &\bf  3.0E-06\\
\hline
1 & 0.5 & 1.4E-01 & 1.5E-01 & 1.2E-01 & 3.4E-01 & 3.5E-01 & 3.3E-01\\
2 & 0.5 & 8.8E-03 & 4.2E-03 & 2.9E-03 & 2.3E-02 & 1.2E-02 & 1.2E-02\\
3 & 0.5 & 7.0E-04 & 2.7E-04 & 2.1E-04 & 2.4E-03 & 1.2E-03 & 1.2E-03\\
\bf 4 &\bf  0.5 &\bf  4.5E-05 &\bf  1.4E-05 &\bf  1.4E-05 &\em  2.0E-04 &\bf  9.1E-05 &\bf  9.0E-05\\
\bf 5 &\bf  0.5 &\bf  2.2E-06 &\bf  6.3E-07 &\bf  6.9E-07 &\bf  1.3E-05 &\bf  5.9E-06 &\bf  5.8E-06\\
\hline
\end{tabular}
\end{table}

\begin{table}[!ht]
\scriptsize
    \centering
\begin{tabular}{|l|l|l|l|l|l|l|l|} 
\hline
$N$        & Interval      & Mean $a_{31}$          & Mean $a_{32}$          & Mean $a_{33}$          & Mean $r_1$        & Mean $r_2$        & Mean $r_3$            \\ 
\hline
\hline
1 & 0.05 & 1.0E-01 & 1.0E-01 & 1.0E-01 & 1.0E-01 & 2.7E-04 & 1.0E-01 \\
\bf 2 &\bf  0.05 &\bf   1.8E-05 &\em  1.0E-04 &\bf  3.0E-05 &\bf  2.0E-06 &\bf  3.5E-06 &\bf  3.6E-05 \\
\bf 3 &\bf  0.05 &\bf  2.8E-07 &\bf  2.0E-06 &\bf  6.0E-07 &\bf  1.7E-08 &\bf  3.9E-08 &\bf  7.9E-07 \\
\bf 4 &\bf  0.05 &\bf  3.5E-09 &\bf  3.2E-08 &\bf  9.6E-09 &\bf  1.3E-10 &\bf  3.1E-10 &\bf  1.3E-08 \\
\bf 5 &\bf  0.05 &\bf  3.8E-11 &\bf  4.4E-10 &\bf  1.3E-10 &\bf  8.0E-13 &\bf  1.9E-12 &\bf  1.8E-10 \\
\hline
1 & 0.1 &  2.0E-01 & 2.1E-01 & 2.1E-01 & 2.0E-01 & 1.0E-01 & 2.0E-01\\
2 & 0.1 &  2.0E-01 & 2.0E-01 & 2.0E-01 & 5.5E-05 & 2.2E-04 & 2.0E-01 \\
\bf 3 &\bf  0.1 &\bf   2.3E-05 &\em  4.9E-04 &\bf  3.5E-04 &\bf  1.3E-06 &\bf  6.7E-06 &\bf  6.0E-05 \\
\bf 4 &\bf  0.1 &\em 1.0E-01 &\em 1.0E-01 &\em 1.0E-01 &\bf  2.1E-08 &\bf  1.4E-07 &\em 1.0E-01\\
\bf 5 &\bf  0.1 &\bf   1.7E-08 &\bf  3.6E-07 &\bf  2.6E-07 &\bf  2.6E-10 &\bf  2.6E-09 &\bf  4.8E-08\\
\hline
1 & 0.2 &  2.1E-01 & 2.1E-01 & 2.1E-01 & 2.1E-03 & 1.0E-01 & 2.0E-01\\
 2 & 0.2 &  3.3E-04 & 1.5E-03 & 1.3E-03 &  4.7E-05 & 3.4E-04 &  2.7E-04\\
\bf 3 & 0.2 &\bf   1.2E-05 &\bf  4.5E-05 &\bf  3.8E-05 &\bf  8.8E-07 &\bf  2.1E-05 &\bf  7.5E-06\\
\bf 4 & 0.2 &\bf   3.4E-07 &\bf  1.2E-06 &\bf  8.5E-07 &\bf  1.3E-08 &\bf  9.4E-07 &\bf  2.0E-07\\
\bf 5 & 0.2 &\bf   8.4E-09 &\bf  2.6E-08 &\bf  1.7E-08 &\bf  1.5E-10 &\bf  3.3E-08 &\bf  4.3E-09\\
\hline
1 & 0.25 & 1.1E-01 & 1.3E-01 & 1.1E-01 & 1.0E-01 & 2.0E-01 & 1.1E-01\\
2 & 0.25 & 2.0E-01 & 2.0E-01 & 2.0E-01 & 1.0E-01 & 2.0E-01 & 2.0E-01\\
 3 &  0.25 &  1.8E-04 & 1.5E-03 &  8.2E-04 &  2.1E-05 &  3.2E-04 &  4.2E-04\\
\bf 4 &\bf  0.25 &\bf  1.0E-05 &\bf  7.2E-05 &\bf  3.6E-05 &\bf  4.9E-07 &\bf  1.5E-05 &\bf  3.2E-05\\
\bf 5 &\bf  0.25 &\bf   5.7E-07 &\bf  3.5E-06 &\bf  1.6E-06 &\bf  8.7E-09 &\bf  5.3E-07 &\bf  2.1E-06\\
\hline
1 & 0.5 &  3.4E-01 & 3.6E-01 & 3.6E-01 & 1.2E-01 & 3.3E-01 & 3.4E-01\\
2 & 0.5 &  3.1E-01 & 3.1E-01 & 3.1E-01 & 2.1E-03 & 8.3E-03 & 3.1E-01\\
3 & 0.5 & 2.0E-01 & 2.0E-01 & 2.0E-01 & 1.5E-04 & 7.9E-04 & 2.0E-01\\
\bf 4 &\bf  0.5 &\em  1.0E-03 &\em 3.2E-03 &\em 3.4E-03 &\bf  9.1E-06 &\bf  6.4E-05 &\em 2.3E-03\\
\bf  5 &\bf  0.5 &\em   1.4E-04 &\em  4.6E-04 &\em  4.7E-04 &\bf   4.6E-07 &\bf   4.3E-06 &\em  3.3E-04\\
\hline
\end{tabular}
\caption{Numerical data representing {\bf mean} relative error in the approximated system to the actual parameter values in a 3-dimensional discrete Lotka-Volterra model. The first column $N$ is the degree (number of terms) of the Taylor expansion. Rows are bolded when the algorithm finds most parameter values up to 4 significant digits, on average, with italics for exceptions. Exceptions are usually narrowly one significant digit less.}
    \label{tab:compmodeldim3mean}
\end{table}
\newpage
\begin{table}[!ht]

    \centering
    \scriptsize
    
\centering
\begin{tabular}{|l|l|l|l|l|l|l|l|} 
\hline
$N$        & Interval      & Med $a_{31}$        & Med $a_{32}$           & Med $a_{33}$           & Med $r_1$         & Med $r_2$         & Med $r_3$             \\ 
\hline
\hline
1 &0.05 &  2.0E-04 &  4.9E-04 &  3.5E-04 &  2.1E-04 &  5.0E-04 &  2.9E-04\\
\bf 2 &\bf  0.05 &\bf   2.1E-06 &\bf  4.3E-06 &\bf  1.4E-06 &\bf  2.0E-06 &\bf  5.5E-06 &\bf  3.2E-06\\
\bf 3 &\bf  0.05 &\bf  9.5E-09 &\bf  3.4E-08 &\bf  8.7E-09 &\bf  1.3E-08 &\bf  3.6E-08 &\bf  2.0E-08\\
\bf 4 &\bf  0.05 &\bf  3.1E-11 &\bf  1.6E-10 &\bf  2.8E-11 &\bf  7.1E-11 &\bf  1.7E-10 &\bf  1.0E-10\\
\bf 5 &\bf  0.05 &\bf  7.3E-14 &\bf  6.7E-13 &\bf  9.9E-14 &\bf  3.2E-13 &\bf  6.5E-13 &\bf  3.2E-13\\
\hline
1 & 0.1 &   1.3E-03 & 9.1E-03 & 3.2E-03 & 1.4E-03 & 6.4E-03 & 3.4E-03\\
\bf 2 &\bf  0.1 &\bf    3.5E-05 &\bf  4.2E-05 &\bf  3.9E-05 &\bf  2.0E-05 &\bf  6.0E-05 &\bf  4.7E-05\\
\bf 3 &\bf  0.1 &\bf    3.6E-07 &\bf  4.8E-07 &\bf  3.2E-07 &\bf  2.1E-07 &\bf  8.6E-07 &\bf  5.7E-07\\
\bf 4 &\bf  0.1 &\bf  3.9E-09 &\bf  4.8E-09 &\bf  3.0E-09 &\bf  1.8E-09 &\bf  8.9E-09 &\bf  7.2E-09\\
\bf 5 &\bf  0.1 &\bf   4.4E-11 &\bf  4.9E-11 &\bf  2.2E-11 &\bf  1.2E-11 &\bf  4.6E-11 &\bf  3.7E-11\\
\hline
1 & 0.2 &  3.1E-03 & 7.7E-03 & 8.7E-03 & 8.1E-03 & 6.1E-03 & 4.0E-03\\
 2 &  0.2 &   8.8E-05 &  3.9E-04 &  3.2E-04 &  1.7E-04 & 4.6E-04 & 6.5E-04\\
\bf 3 &\bf  0.2 &\bf  1.1E-06 &\bf  8.4E-06 &\bf  6.5E-06 &\bf  4.4E-06 &\bf  2.2E-05 &\bf  1.8E-05 \\
\bf 4 &\bf  0.2 &\bf  1.3E-08 &\bf  1.1E-07 &\bf  8.5E-08 &\bf  1.1E-07 &\bf  3.6E-07 &\bf  2.4E-07\\
\bf 5 &\bf  0.2 &\bf  1.2E-10 &\bf  1.1E-09 &\bf  8.5E-10 &\bf  1.7E-09 &\bf  4.2E-09 &\bf  4.9E-09\\
\hline
1 & 0.25 & 4.2E-03 & 1.4E-02 & 4.7E-03 & 1.2E-02 & 2.7E-02 & 3.1E-02\\
2 & 0.25 &  9.6E-05 &  2.6E-04 &  1.3E-04 &  6.4E-04 & 3.8E-03 & 1.7E-03\\
\bf 3 &\bf  0.25 &\bf  1.7E-06 &\bf  6.0E-06 &\bf  2.9E-06 &\bf  1.4E-05 &\bf  7.0E-05 &\bf  4.9E-05\\
\bf 4 &\bf  0.25 &\bf  3.0E-08 &\bf  1.2E-07 &\bf  4.7E-08 &\bf  5.5E-07 &\bf  3.1E-06 &\bf  3.0E-06\\
\bf 5 &\bf  0.25 &\bf  4.3E-10 &\bf  1.7E-09 &\bf  6.1E-10 &\bf  1.7E-08 &\bf  1.0E-07 &\bf  9.6E-08\\
\hline
1 & 0.5 & 2.4E-02 & 3.1E-02 & 2.2E-02 & 7.3E-02 & 8.4E-02 & 5.8E-02\\
2 & 0.5 &  2.4E-03 & 2.4E-03 & 6.1E-04 & 4.0E-03 & 1.1E-02 & 4.9E-03\\
3 & 0.5\bf  &  8.6E-05 & 6.8E-05 & 1.9E-05 &  2.7E-04 & 8.6E-04 &  4.1E-04\\
\bf 4 &\bf  0.5\bf  &\bf  2.2E-06 &\bf  1.8E-06 &\bf  1.2E-06 &\bf  2.0E-05 &\bf  4.5E-05 &\bf  2.1E-05\\
\bf 5 &\bf  0.5\bf  &\bf  4.6E-08 &\bf  2.4E-08 &\bf  2.3E-08 &\bf  8.6E-07 &\bf  1.5E-06 &\bf  8.1E-07\\
\hline
\end{tabular}
\end{table}\begin{table}[!ht]
\scriptsize
    \centering
\begin{tabular}{|l|l|l|l|l|l|l|l|} 
\hline
$N$        & Interval      & Med $a_{31}$        & Med $a_{32}$           & Med $a_{33}$           & Med $r_1$         & Med $r_2$         & Med $r_3$             \\ 
\hline
\hline
 1 & 0.05 &  9.3E-04 & 2.7E-03 &  9.5E-04 &  2.4E-04 &  2.1E-04 &  7.1E-04\\
\bf 2 &\bf  0.05 &\bf  1.2E-05 &\bf  8.5E-05 &\bf  1.6E-05 &\bf  1.4E-06 &\bf  1.5E-06 &\bf  2.8E-05\\
\bf 3 &\bf  0.05 &\bf  1.2E-07 &\bf  1.6E-06 &\bf  6.5E-07 &\bf  5.3E-09 &\bf  1.0E-08 &\bf  4.9E-07\\
\bf 4 &\bf  0.05 &\bf  8.2E-10 &\bf  2.0E-08 &\bf  8.5E-09 &\bf  3.6E-11 &\bf  5.6E-11 &\bf  5.0E-09\\
\bf 5 &\bf  0.05 &\bf  4.6E-12 &\bf  1.8E-10 &\bf  8.6E-11 &\bf  1.4E-13 &\bf  2.2E-13 &\bf  4.2E-11\\
\hline
1 & 0.1 &   3.2E-03 & 2.2E-02 & 9.2E-03 & 7.3E-04 & 1.3E-03 & 4.2E-03\\
\bf 2 &\bf  0.1 &\bf    5.1E-05 &\bf  7.8E-04 &\bf  5.1E-04 &\bf  7.8E-06 &\bf  1.1E-05 &\bf  1.3E-04\\
\bf 3 &\bf  0.1 &\bf  6.2E-07 &\bf  1.2E-05 &\bf  8.2E-06 &\bf  8.0E-08 &\bf  6.8E-08 &\bf  2.5E-06\\
\bf 4 &\bf  0.1 &\bf  5.7E-09 &\bf  1.2E-07 &\bf  8.6E-08 &\bf  7.8E-10 &\bf  1.3E-09 &\bf  2.7E-08\\
\bf 5 &\bf  0.1 &\bf   4.1E-11 &\bf  1.0E-09 &\bf  7.2E-10 &\bf  6.6E-12 &\bf  9.5E-12 &\bf  2.1E-10\\
\hline
1 & 0.2 & 9.0E-03 & 1.5E-02 & 1.5E-02 & 1.2E-03 & 1.6E-03 & 3.3E-03\\
 2 & 0.2 & 2.6E-04 & 7.4E-04 & 7.7E-04 & 2.0E-05 & 1.7E-04 & 2.1E-04\\
\bf 3 &\bf  0.2 &\bf  5.8E-06 &\bf  2.5E-05 &\bf  2.5E-05 &\bf  3.8E-07 &\bf  5.1E-06 &\bf  5.9E-06\\
\bf 4 &\bf  0.2 &\bf  1.1E-07 &\bf  8.2E-07 &\bf  4.7E-07 &\bf  4.8E-09 &\bf  1.0E-07 &\bf  2.2E-07\\
\bf 5 &\bf  0.2 &\bf  2.1E-09 &\bf  1.8E-08 &\bf  8.8E-09 &\bf  4.4E-11 &\bf  1.6E-09 &\bf  3.7E-09\\
\hline
1 & 0.25 & 9.4E-03 & 2.3E-02 & 1.7E-02 & 4.1E-03 & 6.6E-03 & 7.5E-03\\
2 & 0.25 & 4.6E-04 & 1.3E-03 & 1.1E-03 & 7.1E-05 & 1.0E-03 & 4.1E-04\\
\bf 3 &\bf  0.25 &\bf  1.8E-05 &\bf  5.3E-05 &\bf  4.2E-05 &\bf  8.7E-07 &\bf  2.6E-05 &\bf  1.7E-05\\
\bf 4 &\bf  0.25 &\bf  4.6E-07 &\bf  1.6E-06 &\bf  1.1E-06 &\bf  1.5E-08 &\bf  8.2E-07 &\bf  4.6E-07\\
\bf 5 &\bf  0.25 &\bf  9.9E-09 &\bf  3.2E-08 &\bf  2.5E-08 &\bf  2.1E-10 &\bf  2.0E-08 &\bf  9.5E-09\\
\hline
1 & 0.5 & 4.4E-02 & 1.4E-01 & 7.2E-02 & 1.6E-02 & 6.6E-02 & 6.7E-02\\
2 & 0.5 & 1.0E-02 & 1.9E-02 & 1.8E-02 & 4.2E-04 & 3.4E-03 & 1.9E-02\\
3 & 0.5 & 9.0E-04 & 1.8E-03 & 1.4E-03 & 1.1E-05 & 2.3E-04 & 2.0E-03\\
\bf 4 &\bf  0.5 &\bf  8.6E-05 &\em  1.7E-04 &\bf  9.8E-05 &\bf  4.0E-07 &\bf  2.0E-05 &\bf  2.0E-04\\
\bf 5 &\bf  0.5 &\bf  6.8E-06 &\bf  1.4E-05 &\bf  5.0E-06 &\bf  7.7E-09 &\bf  7.8E-07 &\bf  1.3E-05\\
\hline
\end{tabular}\caption{Numerical data representing {\bf median} relative error in the approximated system to the actual parameter values in a 3-dimensional discrete Lotka-Volterra model. The first column $N$ is the degree (number of terms) of the Taylor expansion. Rows are bolded when the median outcome of the algorithm was correct up to 4 significant digits on most parameters, with exceptions in italics. There are very few exceptions.}
    \label{tab:compmodeldim3median}
\end{table}

\subsection{Choice of polynomial system solver}
In our implementation, we  did not incorporate polynomial systems with interval coefficients (see Remark~\ref{rem:4}) and chose HomotopyContinuation.jl~\cite{HomotopyContinuation.jl}. Other solvers, for instance, msolve~\cite{msolve}, can be used as well. We chose the former because it performed faster, which is preferable when performing a large volume of experiments.

\subsection{Computing time efficiency}
We did not explicitly include computation time as an experimental measure. However, for the LPA model, all of the experiments combined (5000 in total) took approximately 45 minutes of sequential computing time, approximately $0.5$ seconds per experiment,  on a computer with CPU Intel(R) Xeon(R) CPU E5-2680 v2 @ 2.80GHz. So, for systems of the size of the LPA model, computing time does not seem to be a limiting factor. For the competition model, we testes how our approach scales with the dimension $n$ of the model growing. The number of parameters in this model is $n^2+n$. Using the above terminology, we ran tests for Taylor approximation degrees up to $5$ and used the interval of $0.05$. The timings are in Table~\ref{tab:cm}, which indicate that medium-sized systems can still be handled in reasonable time with higher degree of approximation.
\begin{table}[h]
    \centering
    \begin{tabular}{|c||c|c|c|c|c|}\hline
        n &2&3&4&5&6  \\\hline
         \# parameters & 6&12&20&30&42\\ \hline
         time & 1 min& 1 min& 2.5 min& 39 min& 10 hrs\\\hline
    \end{tabular}
    \caption{Timings for the competition model with Taylor degree up to $5$ and interval $0.05$.}
     \label{tab:cm}
\end{table}
 For $n=7$, so $56$ parameters total, it took $3$ days to compute parameter estimations for Taylor approximation degrees up to $4$ and finish $60\%$ of the computation (in the implementation, since the polynomial system has groups of disjoint parameter sets, it is broken into subsystems, see Section~\ref{sec:disjoint}) for degree $5$ and then crashing with this error: {\tt OverflowError: Cannot compute a start system}. The mixed volumes returned by the homotopy continuation package for the computed $60\%$ of the roots were about 2 million combined. 
 
\section*{Acknowledgments}
This work was partially supported by: the NSF grants CCF-2212460,  DMS-1760448, and DMS-1853650; CUNY grant PSC-CUNY \#66551-00 54; the Melbourne Research Scholarship; grant of Programa Propio I+D+i 2022, Universidad Politécnica de Madrid, and; grant PID2021-124473NB-I00 from the Spanish MICINN. We are grateful to the Courant Institute of Mathematical Sciences for the computing resources and to the referees for their helpful suggestions.

\bibliographystyle{elsarticle-num.bst}
\bibliography{bib}

\end{document}